%% file: heuristic-and-experimental-carbuncle.tex
\documentclass[a4paper,12pt]{scrartcl}

\pdfoutput=1

\usepackage[utf8]{inputenc}

\usepackage{microtype}
\usepackage{amsmath,bm,amsthm,amssymb,graphicx,transparent}
\usepackage{graphicx,color,transparent}
\usepackage{todonotes}
\usepackage{cite}

\usepackage{bera}
\usepackage[charter]{mathdesign}

\newtheorem{theorem}{Theorem}

\newcommand{\VEC}{\bm}
\newcommand{\MAT}{\bm}

\newcommand{\abs}[1]{\lvert #1 \rvert}
\newcommand{\norm}[1]{\lVert #1 \rVert}
\newcommand{\diag}{\text{\textbf{diag}}}
\begin{document}

\title{Heuristical and numerical considerations for the carbuncle
  phenomenon} 

\author{Friedemann Kemm}

\publishers{\rule{0pt}{1pt}\\\textbf{Keywords:} Carbuncle phenomenon,
  High speed flow, Shock instability, High resolution schemes, Riemann
  solver}

\maketitle

\paragraph*{MSC 2010:} 
  76J20,  
  76K05,  
  76L05,  
  76M99,  
  76N15   


\begin{abstract}
  In this study, we investigate the so called carbuncle phenomenon by
  means of numerical experiments and heuristic considerations. We
  identify two main sources for the carbuncle: instability of the 1d
  shock position and low numerical viscosity on shear waves. We also
  describe how higher order stabilizes the 1d shock position and,
  thus, reduces the carbuncle.
\end{abstract}

\section{Introduction}
\label{sec:introduction}

For the simulation of shock structures in multidimensional gas flows
there are essentially two major approaches: shock fitting and shock
capturing. The idea of the first class of schemes is to exactly detect
the shock position and split the computational domain along the shock
line into two (or for more complex structures even more) domains,
leading to an almost perfect reproduction of shocks in the numerical
solution~\cite{Moretti66,Moretti80,Moretti36}. Disadvantages of this
method include the difficulty to deal with shocks which
unexpectedly evolve in the domain and a practical restriction to
certain shock structures. In order to overcome these
restrictions, more and more scientists started to build their
simulations on shock capturing schemes, which are designed to work
without the knowledge of the exact shock position. Thus, the shock is
captured in a (hopefully) thin layer of grid cells. For an
  overview over shock fitting schemes and some recent developments in
  that area, we refer to~\cite{primer,Bonfiglioli}.

A main ingredient of shock capturing schemes are so called Riemann
solvers: numerical (first order) flux functions, which are based on an
approximate solution of the Riemann problem at the cell face. This
class was later on expanded, starting
with~\cite{schiff_numerical_1980,sw}, through introduction of
Flux-Vector-Splitting schemes, which are based on a splitting of the
physical flux function, but in a wider sense can be considered as
Riemann solvers.  More important for our study however, is the
distinction between complete and incomplete Riemann solvers. While the
first are designed to resolve all waves present in a Riemann problem,
the latter will neglect some waves. Prominent examples of complete
Riemann solvers are the schemes by Roe~\cite{roe-orig} and
Osher~\cite{osher-o}. While these solvers are preferable when complex
wave structures as well as entropy and shear waves are expected,
incomplete solvers are known to be robust in situations dominated by
strong shocks. An example is the HLL-solver~\cite{hll} whose
construction is based only on the two outer waves of the Riemann
problem.
In the eighties and nineties, more and more applications were treated
with above mentioned methods for gas dynamics. The methods were
also extended to other hyperbolic conservation laws like shallow water
or compressible magnetohydrodynamics (MHD).
For a detailed discussion of shock
capturing, we refer to the
textbooks~\cite{toro,leveque-alt,leveque,laney,wesseling,godrav1,godrav2}.

In the context of shock capturing, some irregularities were observed:
properties of the discrete solutions which were by no means
representations of physical phenomena. In gas dynamics simulations
unphysical discrete shock structures and even a complete breakdown of
the discrete shock profile could
appear~\cite{peery1988blunt,quirk}. According to its form in blunt
body problems, it was christened \emph{carbuncle phenomenon}.  Since
the seminal paper of Quirk~\cite{quirk}, an immense amount of research
has been conducted on this instability problem.
The origin of the name
comes from the fact that in strongly supersonic flows against an
infinite cylinder simulated on a body-fitted, structured mesh the
middle part of the resulting bow shock degenerates to a
carbuncle-shaped structure. It was conjectured already by
Quirk~\cite{quirk} that this phenomenon is closely related to other
instabilities such as the so-called \emph{odd-even-decoupling}
encountered in straight shocks aligned with the grid.
Unfortunately, the failure is only found in schemes with high
resolution of shear and entropy waves, so called complete Riemann
solvers, which are needed to properly resolve the boundary
  layers and turbulent structures. This category includes for example
the Godunov, Roe, Osher, HLLC and HLLEM schemes~\cite{toro,hllem}.
These schemes are preferable in calculations involving complex wave
structures as well as boundary layers.

The research on the carbuncle was twofold. On the one hand, the
stability of discrete shock profiles was investigated in one as well
as in several space dimensions. On the other hand, a lot of effort was
put into finding cures for the failure of some schemes in numerical
calculations. For example, many cures that were offered, are based on
an indicator that tells the scheme when to switch to an
  incomplete Riemann solver. These indicators need information from
other cell faces, making the numerical flux function non-local.
It was found that even in one space dimension there are some
instabilities of discrete shock profiles: slowly moving shocks produce
small post-shock oscillations~\cite{quirk,aroracarb,jinliu}. But also
in the case of a steady shock, instabilities can be found depending on
the value of the adiabatic coefficient \(\gamma\) as was shown by
Bultelle et~al.~\cite{bultelle98}. However, the connection to two-dimensional
instabilities is still not fully understood.

The two-dimensional instabilites themselves seem to be closely related to each
other. Chauvat et~al.~\cite{chauvat05} show through an ingenious numerical
investigation that the mechanisms driving the odd-even-decoupling and the
carbuncle are closely related.  Dumbser et~al.~\cite{michael-carbuncle}
present a method to test Riemann solvers for their tendency to
odd-even-decoupling. Here, the basic idea is to discretise a steady shock in
space and test the linear stability of the system of ODEs resulting from the
Method of Lines. This allows for all tested solvers to predict whether they
would evolve an instability or not.  There is also a number of experimental
studies of the carbuncle, especially the influence of the underlying numerical
flux function~\cite{kitamura13,tu2014evaluation,kitamura10,kitamura09}, with
the goal to identify the ``optimal Riemann solver''.  Finally,
Elling~\cite{elling_carbuncle_2009} found a connection to physical shock
instabilities.

Most these investigations have in common that they (a) intend to find a single
source for the carbuncle, (b) do not take into account the influence of the
order of the scheme---they usually compare different Riemann solvers in a
scheme with fixed order---, and (c) do not distinguish between the
contribution of entropy or shear waves to the carbuncle. The most
surprising is case (b) since it is well known that in higher order schemes the
carbuncle is much weaker than in first order; for very high orders, it is
essentially absent. The purpose of this paper is to fill these
gaps in research. We want to study the influence of the (1d) stability of the
shock position and the 2d or 3d features such as vorticity separately. In this
course, we also try to separate the influences of entropy and shear waves. But
the main focus (and main novelty) of this study is that we investigate the
influence of the order of the scheme on the stability of the (1d) shock
position.  We will show how increasing the order of the scheme, despite of
lowering the numerical shear viscosity, stabilizes the 1d shock position.

The outline of the paper is as follows:
In Section~\ref{sec:review-theory} we give a short representative
review of some theoretical results. The insight gained by these
results provides us with the guidelines for our numerical experiments.
In Section~\ref{sec:revi-numer-schem}, we give a review of the schemes
 we use in our numerical experiments. 
The numerical test cases are introduced in
Section~\ref{sec:disc-test-cases}. 
The main results are presented in Sections~\ref{sec:infl-one-dimens}
(one-dimensional issues)
and~\ref{sec:infl-two-dimens} (multi-dimensional issues), followed by
some conclusions and directions for further research in Section~\ref{sec:concl-direct-furth}.

\section{Short review of the theory}
\label{sec:review-theory}

There are many papers dedicated to the carbuncle
phenomenon~\cite{quirk,sanders98,bultelle98,moschetta-vorticity,aroracarb,chauvat05,kim03,pandolfi,park-kwon,zaide-diss,zaide_flux,zaide_shock_2011,shallow-carbuncle,vietnam,elling_carbuncle_2009},
however only few of them discuss the origins of the carbuncle from a
theoretical point of view. Here, we give a short representative review
of some theoretical results.

\subsection{Contribution by Bultelle et al.}
\label{sec:contr-bult-et}

Bultelle et~al.~\cite{bultelle98} investigate steady shocks in one space
dimension. A first study of these was done by Barth~\cite{barth89} who found
that for a perfect gas with~\(\gamma=1.4\), flux functions which enforce the
Rankine-Hugoniot condition at discontinuities may have transition states which
are unstable to perturbations when the preshock Mach number is greater than
six. Bultelle et~al.\ go even further. They prove that the Godunov scheme for
strong steady shocks and an adiabatic coefficient \(1 \leq \gamma^{\ast}
\approx 1.62673\), with \(\gamma^{\ast}\) being a root of
\begin{equation*}
  \gamma^4 + 3 \gamma^3 - 21 \gamma^2 + 17 \gamma + 8 = 0\;,
\end{equation*}
can produce unstable shock profiles. They report that in practice,
after a transient regime, the unstable leads to a stable
profile with the intermediate state in a neighbouring
  cell or even one cell further. If in neighbouring grid slices
normal to the shock front, the shock position jumps in a different
direction, say~\(+2\) cells in one and~\(-2\) cells in another slice,
this leads to an unphysical crossflow along the shock. We discuss this
situation in more detail in Section~\ref{sec:inst-shock-posit}.

\subsection{Contribution by Roe and Zaide}
\label{sec:contribution-roe-et}

While Bultelle et al~\cite{bultelle98} discuss steady shocks in
general, Roe and Zaide~\cite{zaide-diss,zaide_flux,zaide_shock_2011}
focus on the long-time behavior. Although they mainly discuss the
standard Roe solver, the discussion relies on the fact that the
solver tries to establish the exact Rankine-Hugoniot condition at
single shocks. Thus, we can expect that their results also apply to
other Riemann solvers with this property, e.\,g.\ Godunov, HLLE
etc. Roe and Zaide investigate the behavior of steady discrete shocks
with one intermediate point in 1d.  For the Euler equations (and also
for the shallow water equations) it is impossible to find for a
non-trivial steady shock with left and right states~\(\VEC q_l\)
and~\(\VEC q_r\), which are related via the Rankine-Hugoniot
condition, a middle state~\(\VEC q_m\) which is related to both states
via the same condition. As a result, the scheme enforces a
middle state which is unphysical. In some situations,~\(\VEC q_m\) is
not even constant in time: the shock, although unsteady, is trapped in
a single grid cell. And even if~\(\VEC q_m\) is steady, the shock
position in the cell is ambiguous. Depending on the conservative
variable used to compute the shock position one gets different
results. Another feature of these shocks is an overshoot in the
momentum, which is also the mass flux, again indicating the deviation
of the discrete solution from real world physics.

From these results we can draw the conclusion that, at least for
Riemann solvers based on the Rankine-Hugoniot condition,
even shocks which remain in the same grid cell are highly
sensitive to perturbations.

\subsection{Contribution by Elling}
\label{sec:contribution-elling}

\begin{figure}
  \centering
    \def\svgwidth{.6\linewidth}
  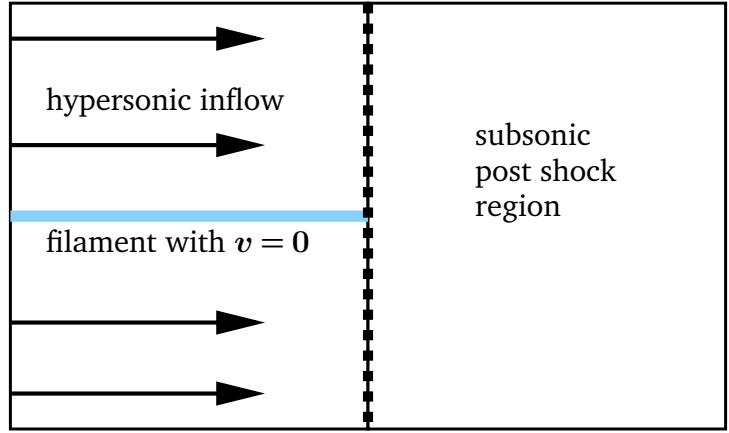
  \caption{Model situation for physical carbuncle.}
  \label{fig:sketchel}
\end{figure}

Elling~\cite{elling_carbuncle_2009} investigates the influence of the
supersonic upstream region on the shock profile. He models the
interaction of a vortex filament with a strong shock. For this
purpose, he starts with a steady shock. In the region upstream of the
shock, he picks the middle slice of the computational grid and
artificially sets the velocity to zero as sketched out
in Figure~\ref{fig:sketchel}. This closely corresponds to
the experimental and theoretical investigations by Kalkhoran and
Smart~\cite{Kalkhoran200063} and by Zhang
et~al.~\cite{zhang-zhang}. It turns out that the carbuncle-like
structure which comes with the Godunov-scheme is similar to the
experimental results. Furthermore, Elling gives numerical evidence
that the structure does not depend on the energy inherent in the
filament. Even more so, the structure is the same for all reasonable
resolutions of the computational domain.
The numerical results obtained with Osher or HLLEM are almost
identical to those obtained with the Godunov scheme in contrast to the
HLLE and other incomplete Riemann solvers. We find that in real world
simulations of a shock interacting with a vortex filament, the
viscosity, especially the shear viscosity, of HLLE and other
incomplete Riemann solvers is too high. It outweighs the physical
viscosity by far. This explains why Elling not only states that the
carbuncle is incurable but also that the carbuncle should not
(completely) be cured, or better: not completely be prevented. The
challenge is to avoid unphysical carbuncles and, at the same time,
allow physical carbuncles. In other words, one has to look for schemes
which allow shear and entropy waves to be well resolved but still
prevent the unphysical breakdown of shock waves.

\section{Review of the numerical schemes for our experiments}
\label{sec:revi-numer-schem}

In order to perform meaningful numerical experiments, one has to
choose a suitable set of numerical methods. Since it is well known
that the choice of the Riemann solver has a strong impact on the
tendency of the scheme to develop a carbuncle, we have to compare
Riemann solvers which differ from each other in some features
coincide in other features. Motivated by the discussion
in the previous section, we choose solvers with different approaches
at shocks: Osher as an example of a solver which abandons the
Rankine-Hugoniot condition as a construction principle and some
members of the Roe/HLL-family which in general force the
Rankine-Hugoniot condition at single shocks\footnote{We will also
  present a modification of HLLEM that yields a behavior at steady
  shocks close to the Osher scheme.}. Since the carbuncle is
characterized by strong shear flows among the Roe/HLL-type solvers,
we select HLLEM, HLLEMCC, and HLLE in order to compare different shear
viscosity mechanisms and their relation to the carbuncle. For the
discussion of the influence of the order of the scheme on the
carbuncle, we employ several standard 1st- and 2nd-order variants as
described below.

\subsection{Basic finite volume scheme}
\label{sec:basic-finite-volume}

For shallow water equations, we employ CLAWPACK~\cite{clawpack}, which is an
implementation of the wave propagation approach. Thus, limiting for higher
order is always done on characteristic variables. Higher order calculations
are done (1) on Cartesian grids with the Superpower
limiter~\cite{limiter,kemm-santiago} and (2) on non-Cartesian grids with the
Albada~3 limiter~\cite{limiter}. Both are smooth limiters. While on Cartesian
grids the more compressive Superpower limiter could be applied, on
non-Cartesian grids one has to go back to a less compressive limiter. In
Section~\pageref{sec:infl-order-scheme} we will discuss the reason why it is
advantageous to opt for a more compressive limiter. The Superpower limiter is
a generalization of the Power limiter by Serna and Marquina~\cite{power}. In
contrast to the original Power limiter, the powers are not fixed but adapted
to the local CFL-number in such a manner that the resulting limiter is always
TVD and approaches third order behavior in smooth regions. The Albada~p
limiters are derived from the van-Albada limiter in the same way as Serna and
Marquina derived the Power limiters from the van~Leer limiter. While Power~2
is just the original van~Leer limiter, Albada~2 is just the original
van~Albada limiter. Albada~3 is the most compressive version that still makes
up a CFL-number independent TVD limiter.

For the Euler results, we employed Euler2d, a 2d-Cartesian code
developed in the Group of Claus-Dieter Munz at Stuttgart
University. The code implements standard finite volumes. For higher
order, direction-wise geometric limiting with minmod on primitive
variables is used.

\subsection{Roe, HLL and their relatives}
\label{sec:hll-its-relatives}

Since the Roe scheme and the HLL-type schemes are closely
related~\cite{habil-kemm,hllem}, we treat them as one family of
schemes. For a better understanding it is convenient to start with the
Roe scheme. 

\subsubsection{Roe}
\label{sec:roe}

The core of the Roe scheme is a consistent local
linearization\cite{roe-orig}, which was first announced by Roe and
Baines~\cite{roe-proc-82} as a means to employ their
TVD-limiters\index{limiter} in a wave-wise manner.
In the linear case,
one finds for the flux function\index{flux!numerical}
\begin{equation}
  \label{eq:43}
  \VEC f (\VEC q) = \MAT A\,\VEC q
\end{equation}
and, thus, for the flux difference\index{flux!difference}
\begin{equation}
  \label{eq:44}
  \VEC f (\VEC q_r) - \VEC f (\VEC q_l) = \MAT A\,(\VEC q_r - \VEC
  q_l)\;. 
\end{equation}
For a local linearization to ensure local conservation, it is
crucial to satisfy a similar condition. Furthermore, the linearized
system should be hyperbolic, and the system matrix should depend
continuously on the left and right states~\(\VEC q_l\) and~\(\VEC
q_r\). Therefore, Roe came up with the following conditions for a
consistent linearization with system matrix~\(\tilde {\MAT A} (\VEC q_l,
\VEC q_r)\):\index{Roe!conditions|textbf}
\begin{gather}
  \label{eq:45}
  \VEC f (\VEC q_r) - \VEC f (\VEC q_l) = \tilde{\MAT A} (\VEC q_l, \VEC
  q_r)\,(\VEC q_r - \VEC q_l)\;,  \\
  \label{eq:47}
  \tilde{\MAT A} (\VEC q_l, \VEC q_r) \to \MAT A(\VEC q) \qquad
  \text{for}\quad (\VEC q_l, \VEC q_r) \to (\VEC q, \VEC q)\;,
   \\ 
  \label{eq:51}
  \tilde{\MAT A} (\VEC q_l, \VEC q_r)\ \text{is diagonalizable for
    all}\ \VEC q_l, \VEC q_r\;.  
\end{gather}
A matrix~\(\tilde {\MAT A} (\VEC q_l, \VEC q_r)\) that satisfies these
conditions is called a Roe matrix or a consistent local linearization
for the underlying system of conservation laws. If there exists a
single state~\(\tilde{\VEC q}=\tilde{\VEC q}(\VEC q_l, \VEC q_r)\)
with
\begin{equation}
  \label{eq:46}
  \tilde{\MAT A} (\VEC q_l, \VEC q_r) = \MAT A(\tilde{\VEC q})\;, 
\end{equation}
then it is called a Roe mean value for \(\VEC q_l, \VEC q_r\).
For the Euler equations of gas dynamics a Roe mean value is given by
\begin{equation}
  \label{eq:108}
  \begin{split}
    \tilde \rho & = \sqrt{\rho_l \rho_r}\;, \\
    \tilde u & = \frac{\sqrt{\rho_l} u_l + \sqrt{\rho_r}
      u_r}{\sqrt{\rho_l} + \sqrt{\rho_r}}\;, \\ 
    \tilde v & = \frac{\sqrt{\rho_l} v_l + \sqrt{\rho_r}
      v_r}{\sqrt{\rho_l} + \sqrt{\rho_r}}\;, \\ 
    \tilde w & = \frac{\sqrt{\rho_l} w_l + \sqrt{\rho_r}
      w_r}{\sqrt{\rho_l} + \sqrt{\rho_r}}\;, \\ 
    \tilde H & = \frac{\sqrt{\rho_l} H_l + \sqrt{\rho_r}
      H_r}{\sqrt{\rho_l} + \sqrt{\rho_r}}\;, \\ 
    \tilde c & = \sqrt{(\gamma - 1)\bigl(\tilde H -
      \tfrac{1}{2}\tilde{\VEC v}^2\bigr)}
  \end{split}
\end{equation}
with~\(\tilde{\VEC v}^2 = \tilde u^2 + \tilde v^2 + \tilde w^2\) in
the full three-dimensional case. For 2d, we just have to omit the
values for the third velocity component~\(w\). 
Similarly, we find a Roe mean value for the shallow water equations by 
\begin{equation}
  \label{eq:109}
  \begin{split}
    \tilde h & = \frac{1}{2}\,(h_l + h_r)\;, \\
    \tilde u & = \frac{\sqrt{h_l} u_l + \sqrt{h_r}
      u_r}{\sqrt{h_l} + \sqrt{h_r}}\;, \\ 
    \tilde v & = \frac{\sqrt{h_l} v_l + \sqrt{h_r}
      v_r}{\sqrt{h_l} + \sqrt{h_r}}\;.
  \end{split}
\end{equation}
Together with wave-wise application of standard upwind, this results
in the numerical flux function
\begin{equation}
  \label{eq:12}
  \VEC g_\text{Roe} (\VEC q_r, \VEC q_l) = \frac{1}{2} \bigl(\VEC
  f(\VEC q_l) + \VEC f(\VEC q_r)\bigr) - 
  \frac{1}{2}\, \abs{\MAT A(\tilde{\VEC q})} \VEC{\Delta q}\;,
\end{equation}
where the absolute value is applied to the eigenvalues of the
matrix. This numerical flux is usually referred to as the standard Roe
solver and is an example for a complete flux. Note that
the wave-wise application of schemes other than standard upwind would
only affect the eigenvalues of the viscosity matrix. In this study, we
do not use the solver in the form~\eqref{eq:12} but with the so called
Harten-Hyman fix~\cite{harten-hyman}, which slightly increases the
numerical viscosity at sonic rarefaction waves and thus prohibits the
sonic glitch. The sonic glitch would otherwise lead to a
representation of sonic rarefaction waves as rarefaction shocks. It is
also possible to resemble the following HLL-type schemes by simply
modifying the eigenvalues of the viscosity matrix in~\eqref{eq:12}.

\subsubsection{HLLE}
\label{sec:hlle}

As an example of an incomplete flux, we employ HLLE\@.
In\cite{hll}, Harten, Lax, and van Leer present and discuss
  a variety of numerical flux functions, the simplest and most robust
  of which is usually called HLL, a scheme with very low
computational cost\footnote{Their more elaborate solvers
    somehow anticipate HLLC~\cite{hllc}}. Their basic idea is to
start from conservation. First, one estimates~\(S_L \leq 0 \leq S_R\)
for the bounding speeds of the Riemann problem given by left and right
states~\(\VEC q_l\), \(\VEC q_r\).  If one uses conservation for
rectangle~\([S_l,S_r]\times [0,1]\) in space and time, the mean value
of the conserved quantities~\(\VEC q\) in the intermediate states of
the Riemann problem can be computed. From this, by integration
over~\([S_l,0]\times [0,1]\) and~\([0,S_r]\times [0,1]\) and
averaging, one obtains the numerical flux
\begin{equation}
  \label{eq:13}
  \VEC g_{\text{HLL}}(\VEC q_r,\VEC q_l) = \frac{1}{2} \bigl(\VEC
  f(\VEC q_r) + \VEC f(\VEC q_l)\bigr) - \frac{1}{2}\, \dfrac{S_R +
    S_L}{S_R - S_L} \bigl(\VEC f(\VEC q_r) - \VEC f(\VEC q_l)\bigr) +
  \dfrac{S_R S_L}{S_R - S_L} (\VEC q_r - \VEC
  q_l) \;.
\end{equation}

We assume now that~\(\tilde{\MAT A} = \tilde{\MAT A} (\VEC q_l,\VEC
q_r)\) is a consistent local linearization, a so-called Roe matrix,
and thus satisfies condition~\eqref{eq:45}. Then~\eqref{eq:13} can be
rewritten as
\begin{equation}
  \label{eq:89}
  \VEC g_{\text{HLL}}(\VEC q_r,\VEC q_l) = \frac{1}{2} \bigl(\VEC
  f(\VEC q_r) + \VEC f(\VEC q_l)\bigr) - \frac{1}{2}\, \dfrac{S_R +
    S_L}{S_R - S_L}  \tilde{\MAT A} (\VEC q_r - \VEC q_l) +
  \dfrac{S_R S_L}{S_R - S_L} (\VEC q_r - \VEC q_l) \;.
\end{equation}
Hence, the viscosity matrix of the HLL-flux is
\begin{equation}
  \label{eq:90}
  \MAT V = \dfrac{S_R + S_L}{S_R - S_L}  \tilde{\MAT A} - 2\,\dfrac{S_R
    S_L}{S_R - S_L} \MAT I
\end{equation}
and has the same left and right eigenvectors as the Roe
matrix~\(\tilde{\MAT A}\) itself. The eigenvalues and thus the
wave-wise viscosity coefficients are
\begin{equation}
  \label{eq:91}
  \dfrac{S_R + S_L}{S_R - S_L}\, \tilde\lambda_k - 2\,\dfrac{S_R S_L}{S_R -
    S_L}
\end{equation}
with~\(\tilde\lambda_k\) being the eigenvalues of~\(\tilde{\MAT
  A}\). Apparently, the choice of the bounding wave speeds is crucial
for the numerical viscosity. In the literature, many choices of the
wave speeds~\(S_L,S_R\) are given. For an overview, the reader is
referred to~\cite{toro}. Here, we mainly rely on the choice suggested
by Einfeldt~\cite{hllem}: If it is affordable to compute the leftmost
and the rightmost wave speed of a consistent local
linearization~\(\tilde\lambda_1\) and~\(\tilde\lambda_m\), and the
flux function~\(\VEC f\) is convex, then set
\begin{equation}
  \label{eq:94}
  S_L  = \min\{\tilde\lambda_1, \lambda_1 (\VEC q_l), 0\}\;, \qquad
  S_R  = \max\{\tilde\lambda_m, \lambda_m (\VEC q_r), 0\}\;.
\end{equation}
This ensures that for both a single discontinuity and a single
rarefaction wave the estimate of the maximal and minimal wave speed is
sharp. The resulting numerical flux is called \emph{HLLE}. Like the
standard Roe scheme, it tries to force the Rankine-Hugoniot condition
at shocks.

\subsubsection{HLLEM}
\label{sec:hllem}

The first scheme that exploited the relation between Roe and HLL type
schemes is the HLLEM scheme for gas dynamics by
Einfeldt~\cite{hllem}. It is an attempt to formulate the Roe scheme as
correction to HLL\@. There are several advantages: The computational
effort is reduced, the adjustment of the viscosity on the acoustic
waves can be easily applied by choosing the bounding speeds~\(S_L\)
and~\(S_R\), the sonic point glitch can be avoided, and the failure of
the standard Roe scheme for strong rarefaction waves can be healed.

The construction is as follows: For the sake of simplicity, we assume
for the eigenvalues~\(\tilde \lambda_1 \leq \tilde\lambda_2 \leq \dots
\leq \tilde\lambda_m\) of the Roe matrix~\(\tilde{\MAT A}\)
that~\(\tilde \lambda_1 \leq 0 \leq \tilde\lambda_m\). Thus, we can
choose~\(S_L = \tilde \lambda_1\) and~\(S_R = \tilde \lambda_m\).
With this setting, the viscosity matrix of the Roe scheme can be
written as 
\begin{align}
  \label{eq:99}
  \MAT V_\text{Roe} & = \MAT V_\text{HLL} + \frac{S_R S_L}{S_R - S_L}
  \tilde{\MAT R}\MAT K\tilde{\MAT L} \\
  \label{eq:101}
  & = \MAT V_\text{HLL} + \frac{\tilde\lambda_m
    \tilde\lambda_1}{\tilde\lambda_m - \tilde\lambda_1} 
  \tilde{\MAT R}\MAT K\tilde{\MAT L}\\
  \intertext{with the anti-diffusion-matrix}
  \MAT K & = \diag (0,\delta_2,\dots,\delta_{m-1},0)\;.
\end{align}
For the two-dimensional Euler equations (\(m=4\)), we find for the
so-called anti-diffusion-coefficients
\begin{equation}
  \label{eq:100}
  \delta_2 = \delta_3 = 2\, \Bigl( 1 - \dfrac{\abs{\tilde u}}{\tilde c +
  \abs{\tilde u}} \Bigr)\;.
\end{equation}
The special structure \(K\) allows us to express the standard Roe flux
by only using the eigenvectors corresponding to the entropy and shear
wave.  Park and Kwon~\cite{park-kwon} show that, independent of the
choice of \(S_L\) and \(S_R\), HLLEM resolves single contact waves
exactly when adhering to the Roe mean values for the
anti-diffusion-term, i.\,e.\ if we stick to~\eqref{eq:101} instead
of~\eqref{eq:99} as originally suggested by Einfeldt. For
HLLEMCC, our modification of HLLEM as discussed in the next
  section, we nevertheless employ the original Einfeldt
setting~\eqref{eq:99}. The loss in resolution of the scheme is rather
small. An advantage of~\eqref{eq:99} is that it deactivates the
anti-diffusion terms automatically for full upwind, i.\,e.\ if one of
the bounding speeds~\(S_L,S_R\) vanishes.

A special case of HLLEM is obtained if we
set~\(S_R = -S_L = \abs{S_\text{max}}\).
In gas dynamics and shallow water flow, the numerical viscosity of the
resulting scheme coincides with the viscosity of the Rusanov/LLF
(Local Lax Friedrichs)
scheme for nonlinear waves and with the viscosity of the standard Roe
scheme for shear and entropy waves.  In the following, we refer to
that scheme as LLFEM.

\subsubsection{HLLEMCC}
\label{sec:hllemcc}

If in the definition of the HLLEM scheme, we replace the
anti-diffusion coefficients~\(\delta_k\) by~\((1-\phi)\, \delta_k\)
with~\(\phi \in [0,1]\), we can smoothly vary between
HLLEM~(\(\phi=0\)) and HLL~(\(\phi=1\)), i.\,e.\ between a
  complete and an incomplete Riemann solver. This is a technique
which we used for our carbuncle cure of HLLEM (HLLEMCC)
in~\cite{kemm-lyon} for the Euler equations and
in~\cite{shallow-carbuncle} for the shallow water equations.

Since the goal of HLLEMCC is to prevent unphysical shear and entropy
waves, its core is the computation of the residual in the
Rankine-Hugoniot condition for these waves:
\begin{equation}
  \label{eq:113}
  \mathfrak r = \VEC f(\VEC q_r) - \VEC f(\VEC q_l) - \tilde u\, (\VEC
  q_r - \VEC q_l)
\end{equation}
with the Roe mean value for the normal velocity~\(\tilde u\). 
Now, we take as our basic indicator the residual relative to~\(\tilde
c\), respectively its Euclidean norm
\begin{equation}
  \label{eq:117}
  \theta = \frac{1}{\tilde c}\,\norm{\mathfrak r}_2\;,  
\end{equation}
which vanishes for all shear and entropy waves.  By introducing
parameters~\(\alpha,\beta\in (0,1)\), we can now define
\begin{align}
  \label{eq:119}
  \phi(\theta,Fr_u) & = \min\{1,\varepsilon\, \theta \max\{0,
  (1-Fr_u^\alpha)\}\}^\beta \\
  \intertext{for shallow water and}
  \label{eq:120}
  \phi(\theta,M_u) & = \min\{1,(\varepsilon\, \theta)^\beta \max\{0,
  (1-M_u^\alpha)\}\} 
\end{align}
for the Euler equations, where~\(Fr_u\) and~\(M_u\) are the
directional Froude and Mach numbers
perpendicular to the cell face. A major advantage of this approach is
that the modification can be applied to shear and entropy waves
separately. This we will use in Section~\ref{sec:infl-visc-entr} to
somehow measure the influence of both wave types on the carbuncle.

However, we still have to choose the parameters. Here, we adhere to the values
as already published\cite{vietnam,shallow-carbuncle,habil-kemm,kemm-lyon}: For
both shallow water and gas dynamics~\(\alpha = \beta =.33\) turned out to be
a good choice. Furthermore, if not stated otherwise, we use~\(\varepsilon =
1/100\) in the gas dynamics case and for the shallow water
equations~\(\varepsilon = 10^{-3}\).

\subsubsection{Shock fix}
\label{sec:shock-fix}

While HLLEMCC modifies the numerical viscosity of HLLEM on the
linearly degenerate waves, here we present a modification of the
numerical viscosity at shocks. We will restrict to steady shocks in
the Euler equations. Again we employ the residual in the
Rankine-Hugoniot condition but this time on the nonlinear waves:
\begin{equation}
  \label{eq:122}
  \mathfrak r  = \VEC f(\VEC q_r) - \VEC f(\VEC q_l) - (\tilde u -
  \tilde c)(\VEC q_r - \VEC q_l) = \tilde c\,\bigl( \mathcal W_2 +
  \mathcal W_3 + 2 \mathcal W_2\bigr)\;,
\end{equation}
for shocks in the left wave, and for shocks in the right wave
\begin{equation}
  \label{eq:123}
  \mathfrak r  = \VEC f(\VEC q_r) - \VEC f(\VEC q_l) - (\tilde u -
  \tilde c)(\VEC q_r - \VEC q_l) = - \tilde c\,\bigl( 2 \mathcal W_1 +
  \mathcal W_2 + \mathcal W_3\bigr)\;.
\end{equation}
Instead of using the residual in
the Rankine-Hugoniot condition relative to the speed of sound, we
employ the sum of the fluxes as our weighting factor. If we take the
shock indicator
\begin{equation}
  \label{eq:124}
  \theta_s = \varepsilon_s \,\Biggl(1\,-\,\frac{\norm{\VEC f(\VEC q_r)
      - \VEC f(\VEC q_l)}_2}{\norm{\VEC f(\VEC q_r) + \VEC f(\VEC
    q_l)}_2}\Biggr)
\end{equation}
with some small positive parameter~\(\varepsilon_s\), we can modify
the Einfeldt choice of wave speeds~\eqref{eq:94} to
\begin{equation}
  \label{eq:125}
  \begin{split}
    S_L  & = \min\{\theta_s(u_l - c_l) + (1-\theta_s)(\tilde u - \tilde
    c),\; u_l - c_l,\; 0\}\;, \\
  S_R  & = \max\{\theta_s(u_r + c_r) + (1-\theta_s)(\tilde u - \tilde
  c),\; u_r + c_r,\; 0\}
  \end{split}
\end{equation}
resulting in a HLLEM-solver that does not enforce the Rankine-Hugoniot
condition at steady shocks. In fact, the numerical viscosity is
slightly increased as like in the Osher scheme, which we describe next.

\subsection{Osher-Solomon}
\label{sec:osher-solomon}

The Osher-Solomon scheme~\cite{osher-o}, a generalization of the Enquist-Osher
scheme~\cite{eo} to systems of conservation laws, was designed with two major
goals: prevent the sonic glitch and represent physically steady states as
discrete steady states. For that purpose, they first had to abandon the
Rankine-Hugoniot condition at shocks as a design principle. Instead, they use
(generalized) Riemann-invariants. While this would lead to a slightly
increased numerical viscosity at shocks, it also allows for stable
steady state representations of steady shocks. Thus, the scheme is a good
candidate for our purposes: measure the influence of the stability of the
shock position on the carbuncle. The resulting numerical flux function reads
as
\begin{equation}
  \label{eq:128}
    \VEC g_\text{Osher} (\VEC q_l,\VEC q_r)  = \frac{1}{2}\,(\VEC f(\VEC
    q_r)+\VEC f(\VEC q_l)) - 
    \frac{1}{2}\,\int_\Gamma \abs{\MAT A(\VEC q)}\:\mathrm d\VEC q\;,
\end{equation}
where~\(\Gamma\) is a path in the state space which connects the left
and right state via integral curves. Later on, Hemker and
Spekreijse~\cite{osher-p} presented an alternative choice
for~\(\Gamma\), which they claim leads to better results. But, since
in our tests we could not find any difference between the results of
both versions, we just refer to it as Osher-scheme without
differentiating for the integration path in~\eqref{eq:128}.



\section{Discussion of the test cases for the numerical investigation
  of the carbuncle}
\label{sec:disc-test-cases}


In our discussion of the carbuncle and its sources, the choice of test
cases plays an important role. They include classical examples like the
double Mach reflection and the blunt body problem, which gave rise to
the name of the carbuncle~\cite{peery1988blunt}, as well as some
\(1\frac{1}{2}\)-dimensional problems like the colliding flow problem,
the steady shock, and the Quirk test. These tests have in common that
they originate from one-dimensional problems which are artificially
augmented with an additional space dimension. Since in these
physically one-dimensional problems there is no inherent source for
the carbuncle, we have to trigger it by adding some noise to the
initial data. While most of these test cases have in common that
in a physical sense we would not expect to see a carbuncle, the
opposite is observed for the Elling test as drawn from the
considerations in Section~\ref{sec:contribution-elling}.

\subsection{Blunt body problem}
\label{sec:blunt-body-problem}

%
In gas dynamics, a popular test case for the carbuncle is the flow
around a cylindrical
obstacle~\cite{chauvat05,michael-carbuncle,kim03,park-kwon,pandolfi,quirk,sanders98}. In
the shallow water case, a similar test would be the flow around a
cylindrical bridge pier. We chose a pier with radius~\(r=1\). As
computational domain we employ a third of the annulus with inner
radius~\(r=1\) and outer radius~\(R=2\). Since the interesting part of
the flow is the inflow region, we restrict the domain in angular
direction to~\(\frac{2\pi}{3},\frac{4\pi}{3}\). The domain is
discretized with 150 cells in the radial direction and 800 cells in
the angular direction. The initial flow is set to the inflow state
everywhere. At the pier we employ wall boundary conditions, at the
other boundaries first order extrapolation.

Although, in principle, it is possible to do comparisons for blunt
body flow also in the gas dynamics case\cite{habil-kemm}, here we
restrict our investigation to shallow water. Since we want to start
with the initial flow set to the inflow state, we could employ the gas
dynamics version of the Osher scheme only for subsonic inflow. Thus,
in this paper, we only show results for the shallow water
case. For the inflow we choose Froude number~\(F\!r=5\).

\subsection{Double Mach reflection}
\label{sec:double-mach-refl}

\begin{figure}
  \centering
  \includegraphics[width=.9\linewidth]{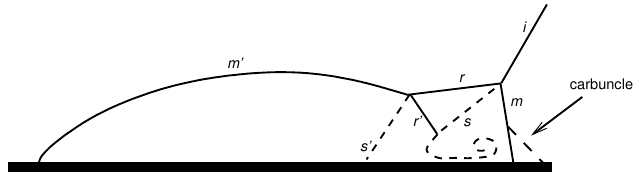}
  \caption{Sketch of Double Mach Reflection with the carbuncle-type
    numerical artifact resulting in a kink in the leading Mach
    stem~\(m\).} 
  \label{fig:dmrsketch}
\end{figure}
A famous test for the quality of a Riemann solver is the Double Mach
reflection problem. It was suggested by Woodward and Colella~\cite{WC}
as a benchmark for Euler codes.  An analytical treatment is found
in~\cite{double-mach},~\cite{ben-dor-buch}, and references therein,
while experimental results are presented
in~\cite{gvozdeva_double_1968} and also
in~\cite[pp.~152~and~168]{ben-dor-buch}. The problem consists of a
shock front that hits a ramp which is inclined by~30~degrees. When the
shock runs up the ramp, a self-similar shock structure with two triple
points evolves. The situation is sketched out in
Figure~\ref{fig:dmrsketch}.  To simplify the graphical representation,
the coordinate system is aligned with the ramp\,--\,as done for the
numerical tests. In the primary triple point, the incident
shock~\(i\), the mach stem~\(m\), and the reflected shock~\(r\)
meet. In the double Mach configuration, the reflected shock breaks up
forming a secondary triple point with the reflected shock~\(r\), a
secondary (bowed) Mach stem~\(m'\), and a secondary reflected
shock~\(r'\).  From both triple points, slip lines emanate. The
reflected shock~\(r'\) hits the primary slip line~\(s\) causing a
curled flow structure.

While the main challenge for a high resolution scheme is to resolve
the secondary slip line~\(s'\)~\cite{dmr,WC}, the main challenge for
first and second order schemes is to correctly represent the leading
Mach stem. For some lower order schemes, the lower end of the Mach
stem moves too fast, leading to a kink in the stem~\cite{quirk}. The
situation is comparable to that for the blunt body problem. According
to the wall boundary conditions, the ramp can be interpreted as the
symmetry-line of a free-stream flow. Since the Mach stem is not
perfectly aligned to the grid lines of a Cartesian grid, the same
mechanisms come into place as for the blunt body problem. Thus, we
expect Riemann solvers which produce a carbuncle for the blunt body
to behave similar in that case. If the flow is split at the
symmetry-line by reinterpreting it again as a wall (or ramp), we have
to expect the lower part of the Mach stem to be kinked as sketched out
in Figure~\ref{fig:dmrsketch}.


For the numerical tests in this paper, we follow the guidelines
in~\cite{dmr}. That means the boundary conditions at the upper boundary
model a slightly smeared shock, and the vertical size of the
computational domain is doubled compared to~\cite{WC}. The only
difference is that, instead of the vertical momentum, we choose the
entropy for our plots. 

\subsection{Colliding flow}
\label{sec:colliding-flow}

This test~\cite[Section~7.7]{astro-leveque} resembles a simplified
model for the starting process of the blunt body test when using the
inflow state as initial data in the complete computational
domain. This is best understood when considering the flow before the
blunt body along the symmetry line. Since the flow is aligned with
that symmetry line, and due to the switch of the sign of the flow
velocity in wall boundary conditions, it behaves essentially like the
left half of a colliding flow in 1d. To turn it into a 2d-test, the
flow is equipped with an additional space direction, in which
everything is expected to be constant. In order to trigger the
carbuncle, the initial state is superimposed with noise that is
generated randomly and has a small amplitude.

For the gas dynamics test, in the initial state, density and pressure
are set to~\(\rho=1,\ p=1\).
The normal velocity is set to~\(u_\text{left/right} = \pm 20\),
the transverse velocity component to~\(v=0\).
To trigger the carbuncle, we superimpose artificial numerical noise of
amplitude \(10^{-6}\)
onto the primitive variables instead of disturbing it in just one
point as was done originally by
LeVeque~\cite[Section~7.7]{astro-leveque}. The computations are done
on~\([0,60]\times[0,30]\) discretized with \(60\times 30\) grid cells.

For shallow water tests, we set the initial height to~\(h=1\) and the
left and right velocities to~\(u=\pm 30\).  The transverse velocity is
zero. Onto this initial state, we superimpose artificial numerical
noise of amplitude \(10^{-6}\), but in this case we add the noise to
the conserved variables. The computations are done
on~\([-2.5,2.5]\times[-2.5,2.5]\) discretized with~\(40\times 40\)
grid cells.

Since the problem is a simple 2d-extension of a one-dimensional problem, the
results are presented in scatter-type plots: we slice the grid in
\(x\)-direction along the cell faces and plot the entropy for all slices at
once.

\subsection{Steady shock}
\label{sec:steady-shock}

While the colliding flow test models the starting process of the blunt
body flow, the steady shock test, introduced by Dumbser
et~al.~\cite{michael-carbuncle}, features a simplified model for the
converged shock in the blunt body flow. Following Dumbser et~al., we
set in the upstream region~\(\rho =1,\ u=1\).
The upstream Mach number is set to~\(M=20\),
the transverse velocity component to~\(v=0\).
The shock is located directly on a cell face.  To trigger the
instability of the discrete shock profile, we add artificial numerical
noise of amplitude \(10^{-6}\)
to the primitive variables in the initial state. The computations are
done on~\([0,100]\times[0,40]\)
discretized with~\(100\times 40\) grid cells.

For shallow water we set the water height and the Froude number at the inflow
to~\(h=1\) and \(Fr=30\) respectively. At the outflow, we simply employ
extrapolation boundary conditions.  Again, we add artificial numerical noise,
this time of amplitude \(10^{-3}\), to the conserved variables in the initial
state. The computations are done on~\([-2.5,2.5]\times[-2.5,2.5]\) discretized
with~\(100\times 40\) grid cells.
%

Again for the presentation of the results, we employ scatter-type
plots as described for the colliding flow problem.

\subsection{Quirk test}
\label{sec:quirk-test}

Quirk~\cite{quirk} introduced a test problem which is known as
Quirk test. Contrary to the preceding example, it is not a
one-dimensional Riemann problem, but consists of a shock running down
a duct. The shock is caused by Dirichlet-type boundary conditions on
the left boundary with~\(\rho=5.26829268\), \( u=4.86111111\),
\(p=29.88095238\), while the flow field is initialized with~\(\rho = 1,\
u=v=0,\ p=1/\gamma\). Originally, a disturbance of the middle grid
line was used to trigger the instability~\cite{quirk}. Because the
computations are done with a Cartesian code, we instead use numerical
noise in the same manner as for the steady shock and the colliding
flow problem. The only difference lies in the amplitude of the
perturbation, here \(10^{-3}\). The computations are done
on~\([0,1600]\times[0,20]\) discretized with \(1600\times 20\) grid
cells. 
In this study, we only perform the test for the Euler equations. For a
similar test in shallow water, we refer to~\cite{vietnam}.

Again we use scatter-type
plots (as described above) to present the results.

\subsection{Elling test}
\label{sec:elling-test}

The Elling test is the experiment which was already described in
Section~\ref{sec:contribution-elling}. 
The
initial condition is a modified version of the
steady shock test, cf.\ Section~\ref{sec:steady-shock}. The
region to the right of the shock remains unchanged. In the supersonic
inflow region, only the middle \(x\)-slice is changed. Here the velocity is
set to zero. This is done to model a vortex layer hitting the shock
front. 
Again, we only perform the test for the Euler equations and refer
to~\cite{vietnam} for a similar test in shallow water.

\section{Influence of one-dimensional issues}
\label{sec:infl-one-dimens}

Although the carbuncle is a multi-dimensional issue, it
is obvious that, especially on Cartesian grids, one-dimensional issues
can drive multi-dimensional effects.  

\subsection{Instability of shock position in first order schemes}
\label{sec:inst-shock-posit}

Two of the instabilities discussed in Section~\ref{sec:review-theory}
are purely one-dimensional. Both are instabilities of the shock: the
instability of the shock position relative to the grid
(Sec.~\ref{sec:contr-bult-et}) and the instability and ambiguity of
the shock position within the grid cell
(Sec.~\ref{sec:contribution-roe-et}).
\begin{figure}
  \centering
    \def\svgwidth{.7\linewidth}
  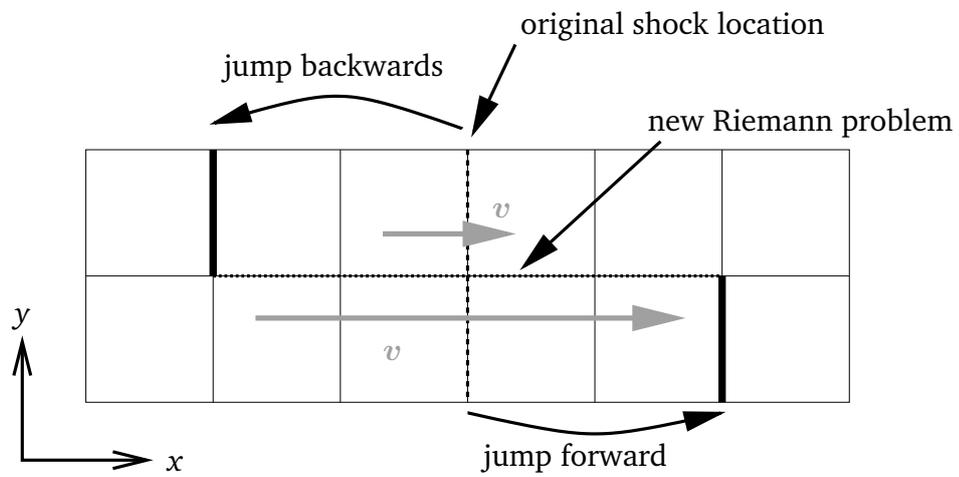
  \caption{Unstable two-dimensional shock profile}
  \label{fig:jumping}
\end{figure}
Figure~\ref{fig:jumping} illustrates how the instability of the
shock position can destroy a discrete shock profile. As Bultelle
et~al.~\cite{bultelle98} point out, the shock position might jump by
up to two grid cells. Figure~\ref{fig:jumping} shows the worst case
scenario: in one grid slice, it jumps two cells upstream, in the
neighboring slice, it jumps two cells downstream. Thus, at a length of
four grid faces, we created a new Riemann problem with all types of
waves~\cite{habil-kemm}. In the depicted situation, a strong flow
downwards would be initiated. The instabilities and the ambiguity
discussed in Section~\ref{sec:contribution-roe-et} can affect the
discrete shock profile in a similar way. Since they are highly
sensitive for perturbations, cross-flow might be induced within the
grid slice containing the original shock itself.

\begin{figure}
  \centering
  \includegraphics[width=\linewidth]{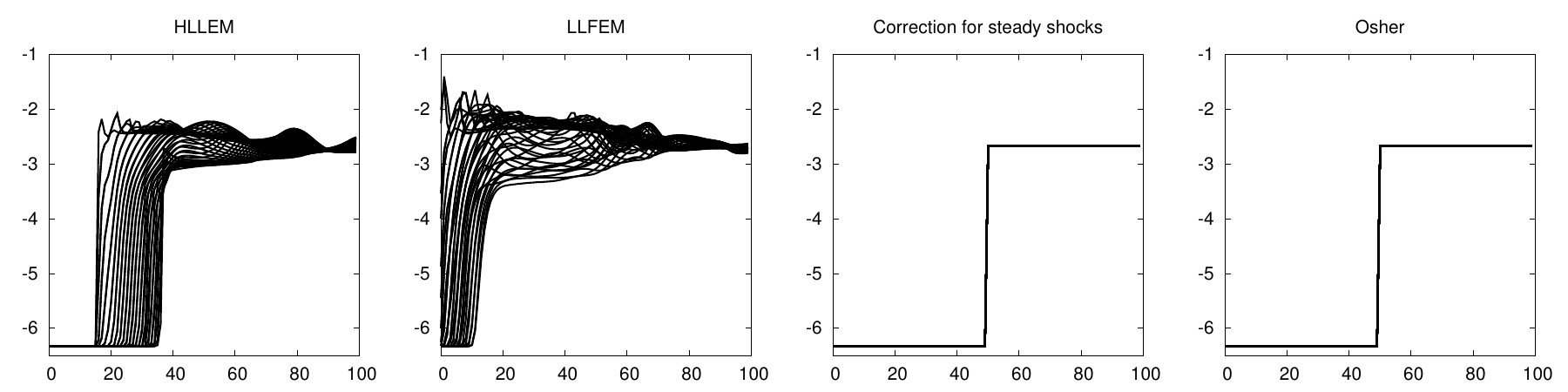}
  \caption{Scatter-type plots of entropy for steady shock with
    different numerical fluxes at~\(t=1000\). }
  \label{fig:steadysteady}
\end{figure}

This situation could be avoided if the 1d shock position was
stable. However, as seen from the discussions in
Sections~\ref{sec:contr-bult-et} and~\ref{sec:contribution-roe-et},
this would mean to abandon the requirement of Riemann solver which
exactly reproduces the Rankine-Hugoniot condition at a single
shock. As mentioned above, the Osher solver replaces this by the
requirement of yielding steady discrete solutions for any steady
discontinuity and employs Riemann invariants over all nonlinear
waves. Thus, the Osher scheme seems to be a good candidate to avoid
the instability of the 1d shock position, much the same as the
HLLEM-scheme with the modification at steady
shocks~(Sec.~\ref{sec:shock-fix}). Another solver which also
disregards the Rankine-Hugoniot condition is LLFEM\@. The difference to
Osher and the steady shock fix is that the numerical viscosity on the
nonlinear waves is much higher.
As the results in Figure~\ref{fig:steadysteady} show, the
LLFEM cannot prevent the carbuncle while the other two
can. From that, we conclude that it is important to add the correct
amount of viscosity in order to stabilize the shock position.

\begin{figure}
  \centering
  \includegraphics[width=\linewidth]{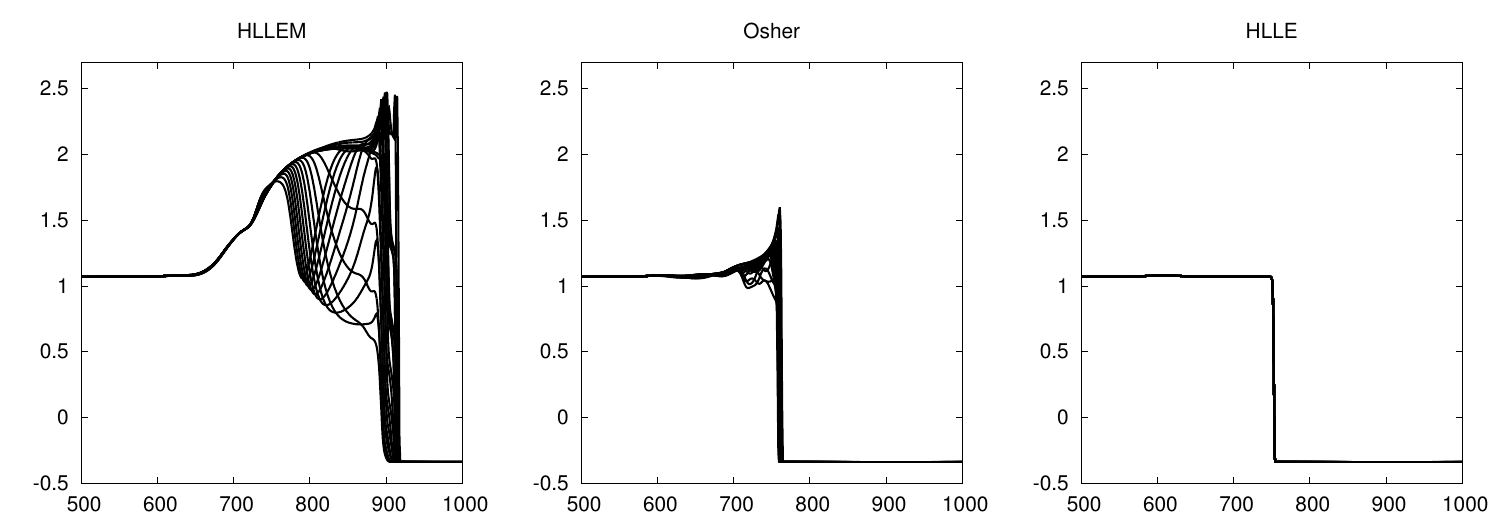}
  \caption{Scatter-type plots of entropy for Quirk test with different
    numerical fluxes at~\(t=125\).}
  \label{fig:oshquirk}
\end{figure}

Figure~\ref{fig:oshquirk} reveals that if we guarantee steady discrete
representations of steady states this is not sufficient to guarantee a
proper representation of unsteady flows. The Quirk test is employed
and shows that the Osher scheme still might produce carbuncle-like
structures, although much weaker than with, e.\,g., HLLEM. 

\begin{figure}
  \centering
  \includegraphics[width=\linewidth]{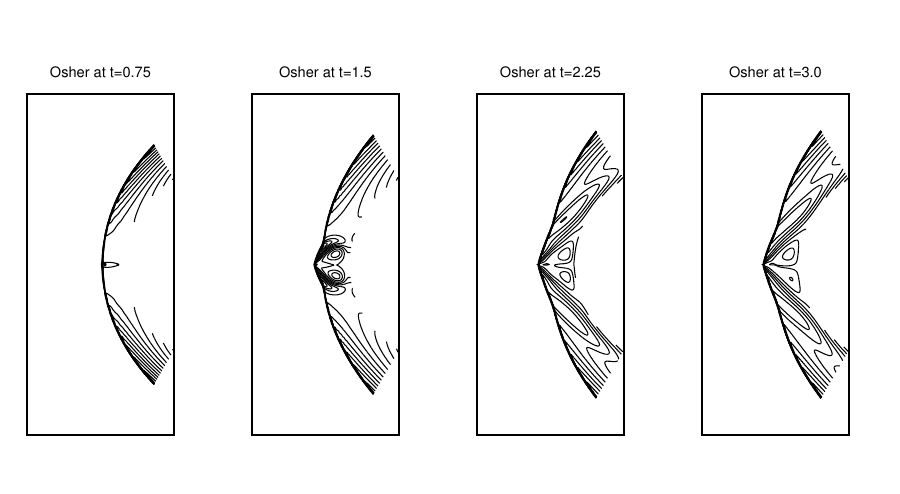}
  \caption{Flow around cylindrical pier in shallow water with 1st
    order Osher scheme at different times. Energy shown.}
  \label{fig:oshpil}
\end{figure}

Our setting for the blunt body problem results in a highly transient
starting phase which eventually passes into a steady state. This
raises the question which property of the Osher scheme would be
dominant: the tendency to produce a carbuncle in a transient flow or
the guarantee for steady discrete representations of steady
flows. Thus, in Figure~\ref{fig:oshpil}, we show the shallow water
flow around a cylindrical pier at different times. At time~\(t=1.5\),
it is easy to see that the scheme produces a slight carbuncle-type
structure. When the discrete flow field eventually becomes steady,
obviously that carbuncle is somehow smoothed out.
The solution is steady but not physical, confirming the statement
by Robinet et~al.~\cite{robinet00} that the carbuncle can lead to unphysical steady states.

\subsection{Influence of the order of the scheme}
\label{sec:infl-order-scheme}

\begin{figure}
  \centering
  \def\svgwidth{.7\linewidth}
  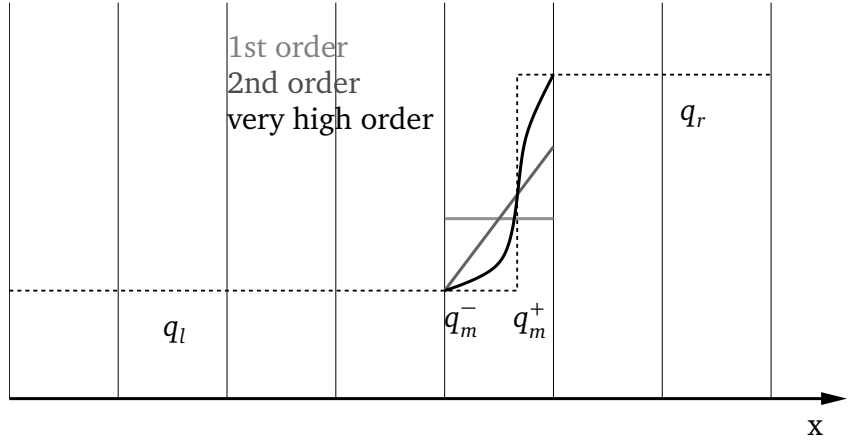
  \caption{Reconstruction of shock profile (dashed line) on finite
    volume grid with different orders}
  \label{fig:shockho}
\end{figure}

Since it is often reported that increasing the order of the scheme applied in
the computation reduces the carbuncle, here we investigate the relation
between the order of the scheme and the 1d-stability of the discretized
shock. At this point, we should stress out that throughout this paper the term
\emph{order} refers to the design order of the scheme, which is only achieved
in smooth parts of the flow field, and not to the actual order of the scheme
which would automatically drop in the vicinity of shocks. As was pointed out
by Roe and Zaide~\cite{zaide-diss,zaide_flux,zaide_shock_2011}, a major role
is played by the fact that it is impossible, at least for gas dynamics and
shallow water flows, to split a shock satisfying the Rankine-Hugoniot
condition into two consecutive shocks which both would satisfy the
Rankine-Hugoniot condition. This situation improves for higher order, at least
when geometric reconstruction is applied. For the computation of the
inter-cell fluxes, two states are available, one at the left cell boundary and
one at the right cell boundary. Thus, for a steady shock it would be
sufficient to satisfy
\begin{equation}
  \label{eq:9}
  \VEC f(\VEC q_l) = \VEC f(\VEC q_m^-) = \VEC f(\VEC q_m^+) = \VEC
  f(\VEC q_r)\;, 
\end{equation}
which can easily be achieved by
\begin{equation}
  \label{eq:10}
  \VEC q_m^- = \VEC q_l\;,\qquad \VEC q_m^+ = \VEC q_r\;. 
\end{equation}
This is (for a scalar situation) sketched out in Figure~\ref{fig:shockho}. 
For the sake of simplicity, in the following, we restrict our
considerations to the scalar case. It is easy to see how the results
can be transferred back to the systems case. 

Figure~\ref{fig:shockho} shows how the situation
improves with increasing order of the scheme. If, for instance, we
employ a polynomial reconstruction in the cell where the shock is
located, for high orders, condition~\eqref{eq:10} can be ensured
for almost all shock positions
without sacrificing monotonicity:

\begin{theorem}\label{theorem1}
  Given the situation depicted in Figure~\ref{fig:shockho}, let the
  cell with the shock be~\([x_{i-1/2},x_{i+1/2}]\)
  and~\(\Delta x = x_{i+1/2} -x_{i-1/2}\).
  Furthermore let~\(\Delta q = q_r - q_l \neq 0\).
  Let the shock be located at~\(x_{i-1/2} + \theta\Delta x\)
  with~\(\theta \in [0,1]\).
  \begin{enumerate}
  \item For any~\(\theta \in [0,1]\),
    a polynomial reconstruction~\(p_{(n)}\)
    with degree less or equal two can be found such that
    condition~\eqref{eq:10} is satisfied. \label{item:1}
  \item In the cell~\([x_{i-1/2},x_{i+1/2}]\),
    the reconstruction can be made monotone for polynomial
    degree~\(\leq n\)
    if~\(\theta \in [\frac{1}{n+1},
    1-\frac{1}{n+1}]\).  \label{item:2}
  \end{enumerate}
\end{theorem}
\begin{proof}
  Statement~\ref{item:1} is obvious and well known. It was already
  used by van~Leer in his work on higher order
  methods~\cite{vanLeer4}.

  For the proof of statement~\ref{item:2}, we can assume, without
  restriction,
  \begin{equation*}
    [x_{i-1/2},x_{i+1/2}] = [0,1]\;,\qquad q_l = 0\;,\quad q_r =
    1\;,\quad\text{and}\quad \theta \geq 1/2\;. 
  \end{equation*}
  All other cases can be derived from this by symmetry, scaling, and
  translation. 

  It is easy to see that with above settings
  \begin{equation}
    \label{eq:14}
    \int_0^1 q(x)\:dx = 1 - \theta\;.
  \end{equation}
  On the other hand, we know for any monomial
  \begin{equation}
    \label{eq:16}
    \int_0^1 x^n\:dx = \frac{1}{n+1}\;.
  \end{equation}
  Furthermore all monomials with degree greater of equal one are
  monotonously increasing in~\([0,1]\)
  and attain the values~\(0\)
  at~\(x=0\)
  and~\(1\)
  at~\(x=1\).
  Obviously, the same is true for all weighted means of such
  polynomials as long as the weights are non-negative. Thus, for
  any~\(\theta\in [\frac{1}{2},\frac{1}{n+1}]\),
  we can find a polynomial~\(p_n\)
  of degree~\(\leq n\), monotonously increasing in~\([0,1]\), with
  \begin{equation}
    \label{eq:17}
    p_n(0) = 0\;,\quad p_n(1) = 1\;,\qquad\text{and}\quad \int_0^1
    p_n(x)\:dx = 1 - \theta\;.
  \end{equation}
  As already mentioned, the general case follows from this by
  symmetry, scaling, and translation.
\end{proof}
Note that the linear reconstruction of case~\ref{item:1} in
Theorem~\ref{theorem1} is not achieved by standard second order
schemes. Due to their restricted stencil, they cannot distinguish
between the situation of Theorem~\ref{theorem1}, case~\ref{item:1}
(slope~\(\Delta q/\Delta x\)
in the middle cell), and a linear state with
slope~\(\Delta q/(2\Delta x)\).
But in order to achieve second order, they have to reconstruct linear
states exactly.  Slope~\(\Delta q/\Delta x\)
in the middle cell can be achieved, e.\,g., by the
limiter~\(\varphi_\text{Sweby}\),
the upper bound of the Sweby region, as described
in~\cite[Section~2.4.2]{limiter}, leading to a first order scheme. For
other~\(\theta\in [-1,1]\)
it would still satisfy one of the identities~\eqref{eq:10}. While
Minmod cannot satisfy any of the identities~\eqref{eq:10}
for~\(\theta\in (-1,1)\),
Superbee shares the behavior of~\(\varphi_\text{Sweby}\)
for~\(\abs{\theta} \geq 1/3\) and the MC-Limiter for~\(\abs{\theta}
\geq 1/2\). 

In this context, it is also worth to note that some third order
schemes, although formally based on linear reconstructions,
e.\,g.~\cite{doppellog,mirojcp} in the final analysis still employ
parabolic reconstructions. For each cell, they compute two linear
reconstructions in the manner described
in~\cite[Section~2.4.2]{limiter}, one ensuring third order for
left-going waves and another one ensuring third order for right-going
waves. While~\(q_m^-\)
is taken from the first reconstruction, \(q_m^+\)
is taken from the latter. Although it is, in general, not possible to
reinterpret this as a single linear reconstruction, it is always
possible to reinterpret it as a parabolic reconstruction. But due to
their restricted stencil, they do, in general, not satisfy the
identities in equation~\eqref{eq:10}. Strangely enough, none of the
authors of such limiters considers the resulting parabolic
reconstruction, and, therefore, none of them checks if this parabolic
reconstruction is indeed monotone.

  Since many modern schemes, like ENO/WENO, do not explicitly enforce
  monotone reconstructions in the cells, for these schemes, we can
  expect identities~\eqref{eq:10} to be satisfied in most cases, and,
  if not so, at least approximated with very small error. Thus, for
  schemes of order greater or equal three, we we might expect
  physically steady shocks to be represented as discrete steady shock,
  independent of the Riemann solver and the position of the shock, and
  hope for a similar behavior for moving shocks.

\begin{figure}
  \centering
  \includegraphics[width=\linewidth]{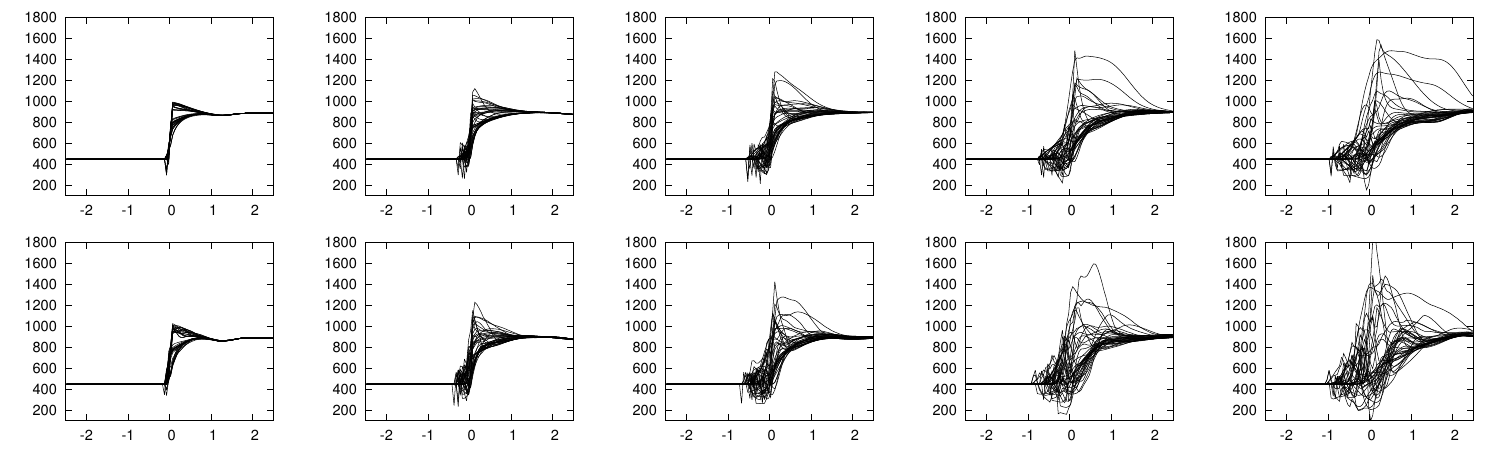}
  \caption{Steady shock problem in shallow water at
    times~\(t=0.2,\,0.4,\,0.6,\,0.8,\,1.0\) with Roe scheme,
    energy shown. Upper row: first order; second row: second order
    with Superpower limiter.}
  \label{fig:steadho}
\end{figure}

\begin{figure}
  \centering
  \includegraphics[width=\linewidth]{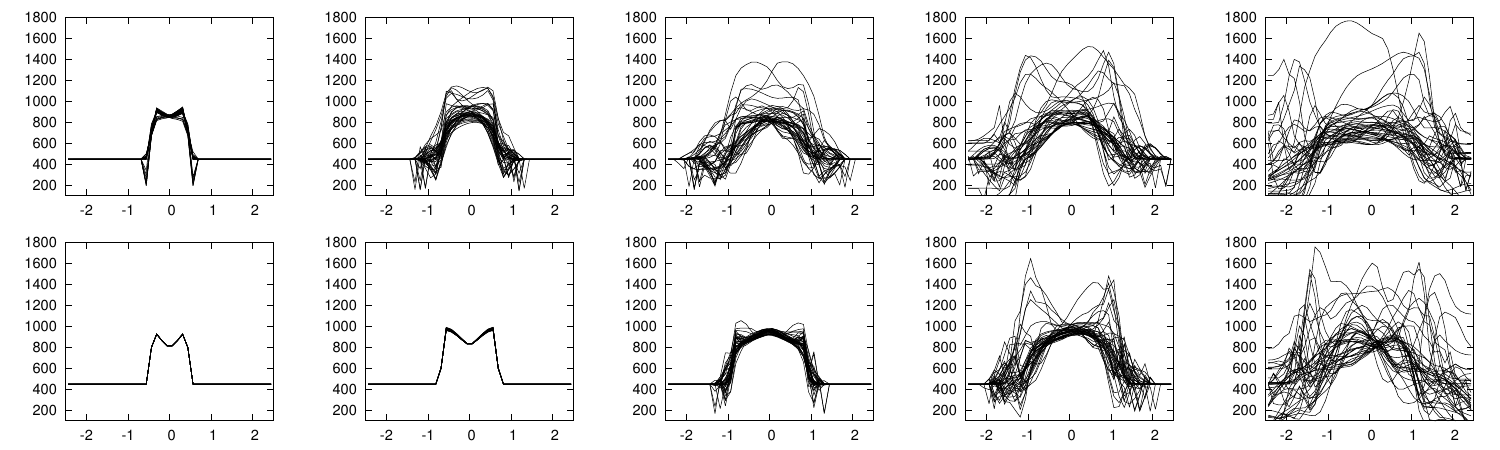}
  \caption{Colliding flow problem in shallow water at
    times~\(t=2/3,\,1,\,4/3,\,5/3,\,2\) with Roe scheme,
    energy shown. Upper row: first order; second row: second order
    with Superpower limiter.}
  \label{fig:collho}
\end{figure}

\begin{figure}
  \centering
  \includegraphics[width=\linewidth]{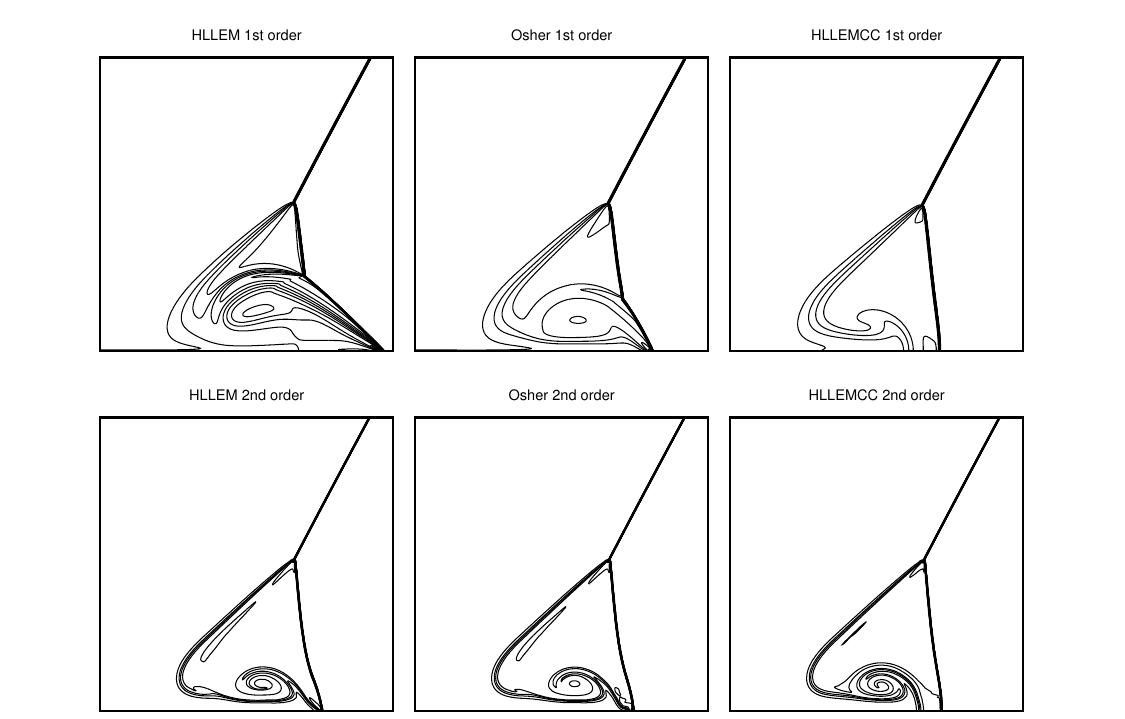}
  \caption{Leading shock structure of double Mach reflection in gas
    dynamics at~\(t=0.2\) with different numerical fluxes, entropy
    shown. Upper row: first order, lower row: second order. }
  \label{fig:dmr15}
\end{figure}

\begin{figure}
  \centering
  \includegraphics[width=\linewidth]{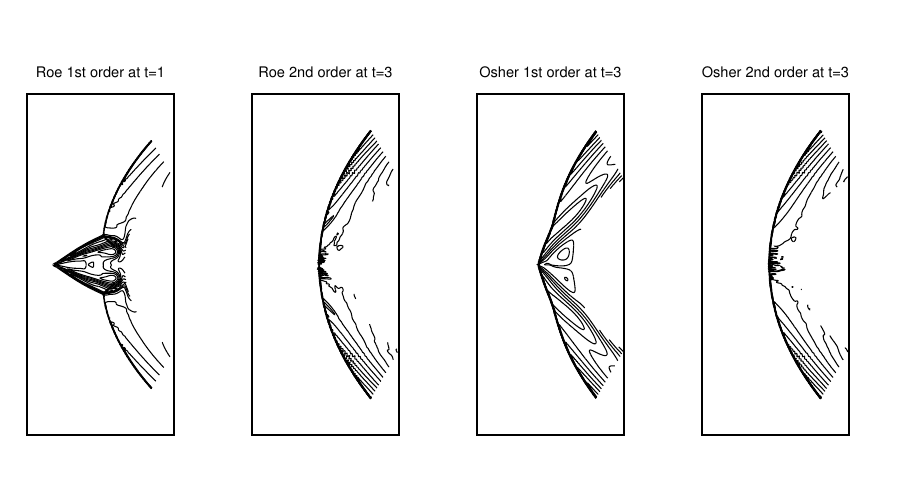}
  \caption{Flow around cylindrical pier in shallow water at Froude
    number~\(Fr=5\) with Roe and Osher, first order and second order
    with Albada~3.}
  \label{fig:pillar15}
\end{figure}

\begin{figure}
  \centering
  \includegraphics[width=\linewidth]{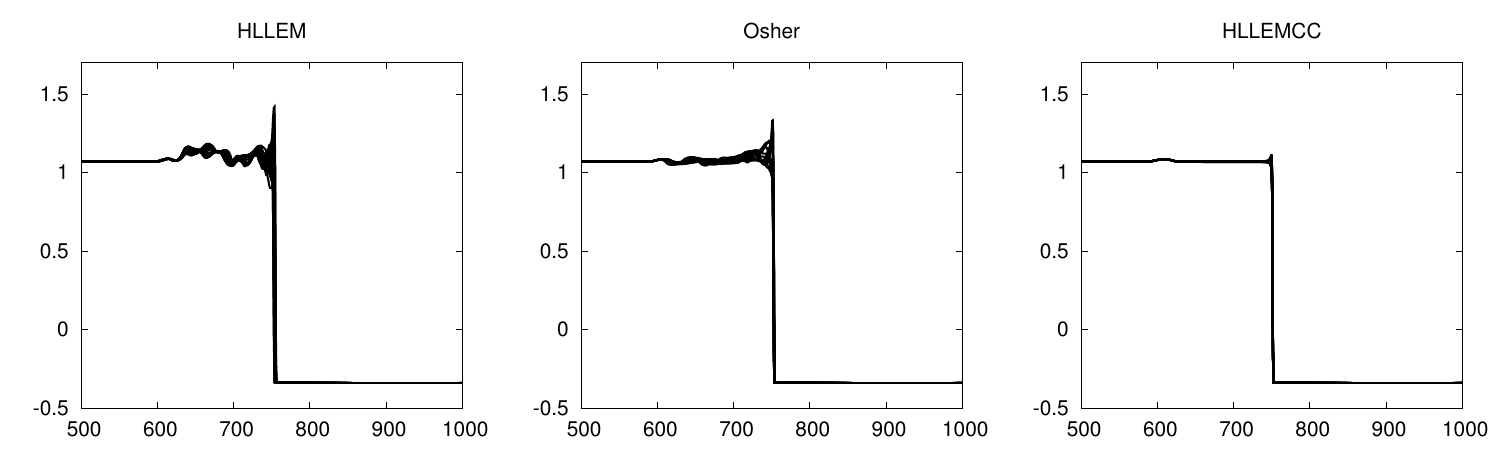}
  \caption{Scatter-type plots of entropy for Quirk test with different
    numerical fluxes and second order at~\(t=125\).}
  \label{fig:o2quirk}
\end{figure}

\section{Influence of two- or three-dimensional issues}
\label{sec:infl-two-dimens}

For years, the research on the carbuncle concentrated on
two-dimensional issues, although the Osher scheme was already in
widespread use, and the carbuncle was known to be rarely found in very
high order schemes or on unstructured grids. Here, we started with
one-dimensional issues to better understand how they, when everything
is generalized to two or three space-dimensions, interact with two-
and three-dimensional issues. Hence, in this section, we want to
concentrate on the interplay of 1d and 2d (or 3d) issues.

\subsection{Numerical shear viscosity}
\label{sec:numer-shear-visc}

At this point, we have to refer the reader back to
Figure~\ref{fig:jumping}. It is obvious that the vertical flow induced
by the instability of the shock position in turn induces a strong
shear flow not only in the newly created Riemann problem, but even
more along the original shock profile. In summary, some kind of
turbulence at the original shock position is created which is
superimposed onto the original flow. Due to its construction, the
HLLEMCC solver distinguishes between shear waves which are
superimposed onto nonlinear waves and shear waves which are not. On
the first, it behaves like HLLE, thus damping the turbulence, on the
latter, it behaves like HLLEM, here allowing, e.\,g., for well
resolved boundary layers. In our previous
works~\cite{vietnam,shallow-carbuncle,habil-kemm,kemm-lyon}, we could
show that HLLEMCC prevents the carbuncle while allowing for good
resolution of physical shear waves.

\begin{figure}
  \centering
  \includegraphics[width=.7\linewidth]{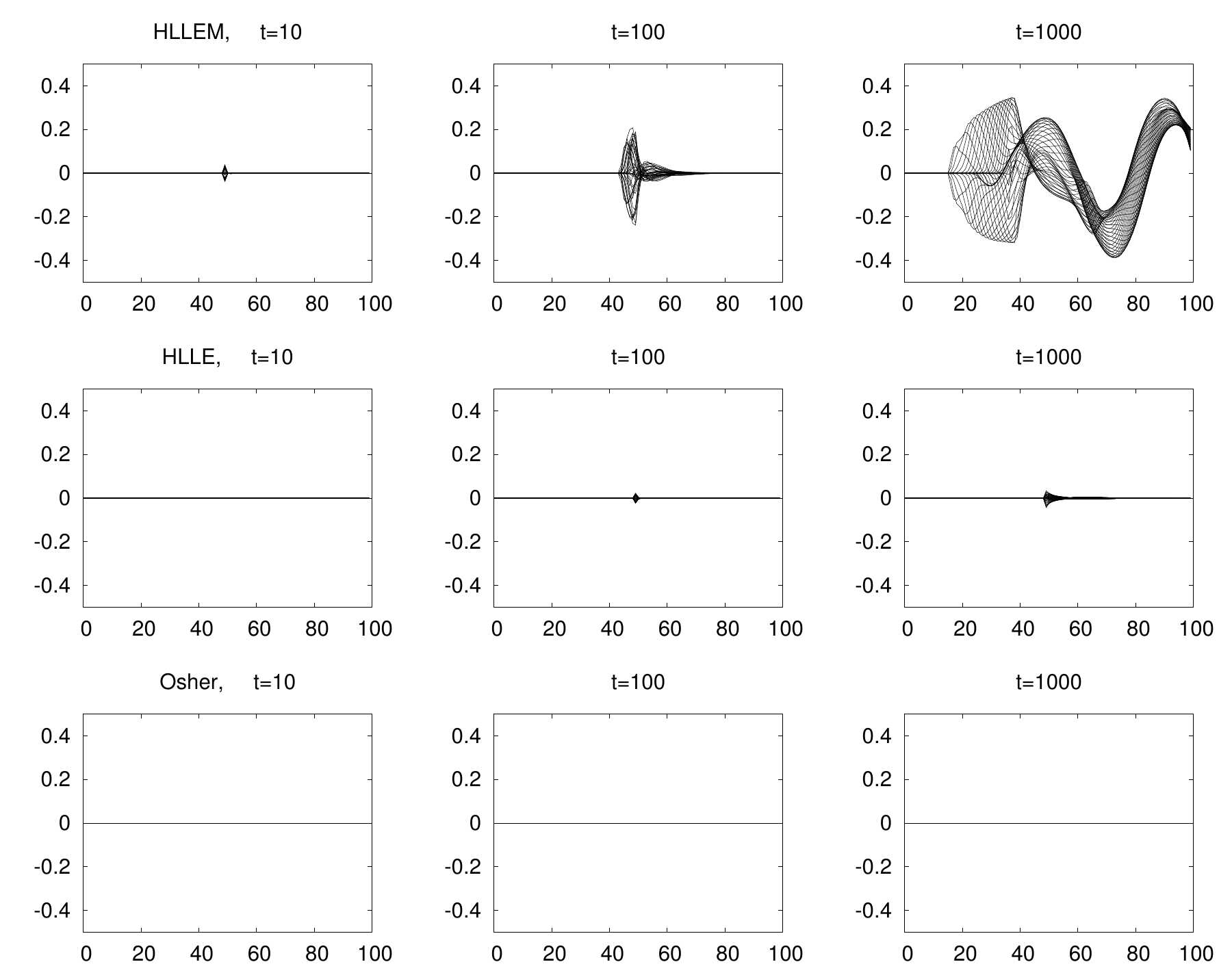}
  \caption{Transverse velocity at different times in gas dynamics
    steady shock problem for different solvers.}
  \label{fig:evo}
\end{figure}
In Figure~\ref{fig:evo}, we demonstrate how the above described
mechanism drives the carbuncle. We show the transverse velocity at
different times for HLLEM, HLLE, and the Osher scheme. By stabilizing
the 1d shock position, the Osher scheme keeps the transverse velocity
at about the magnitude of the artificial numerical noise
introduced in the initial state. Thus, no carbuncle arises. The HLLE scheme,
by its excessive shear viscosity, keeps the transverse velocity below
some threshold and thus also avoids the carbuncle. For HLLEM, there
is no mechanism to damp the turbulence along the original shock. Over
time, a carbuncle evolves.

\subsubsection{Shear viscosity and order of the scheme}
\label{sec:shear-visc-order}

\begin{figure}
  \centering
  \includegraphics[width=\linewidth]{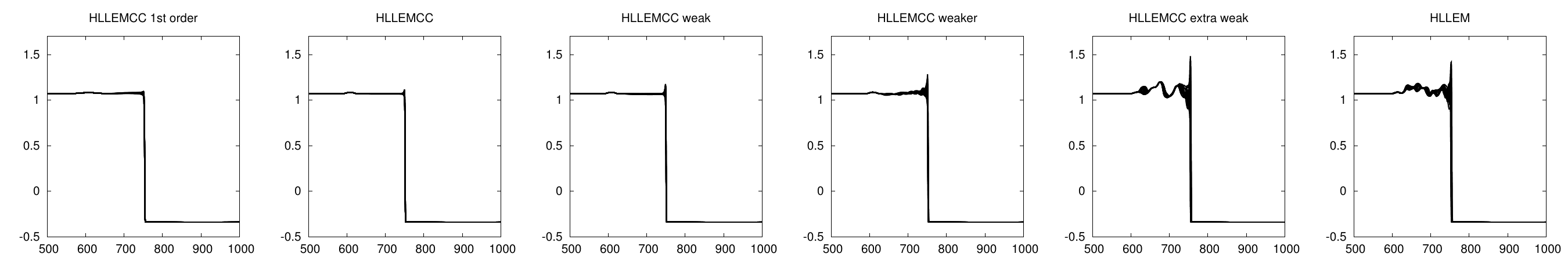}
  \caption{Quirk test at~\(t=125\) with 1st order standard HLLEMCC and different
    versions of 2nd order HLLEMCC scheme.}
  \label{fig:quirkccord2}
\end{figure}

\begin{figure}
  \centering
  \includegraphics[width=\linewidth]{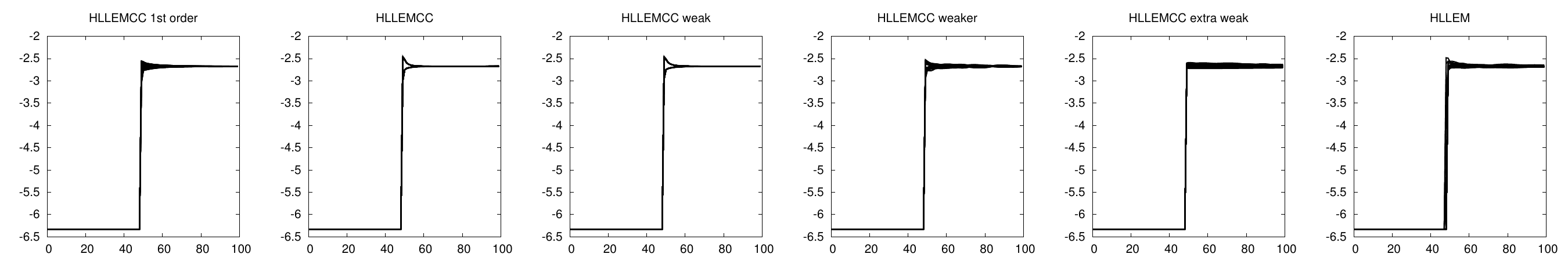}
  \caption{Steady shock in gas dynamics at~\(t=1000\) with 1st order
    standard HLLEMCC and different versions of 2nd order HLLEMCC scheme.}
  \label{fig:steadyccord2}
\end{figure}

As we have seen in Section~\ref{sec:infl-order-scheme}, higher order has a
stabilizing effect on the 1d shock position. This raises the question if for
second order schemes, the carbuncle correction in HLLEMCC might be
relaxed. The answer is not obvious since raising the order also lowers the
viscosity on the shear waves. In Figure~\ref{fig:quirkccord2}, we give a
comparison of first order HLLEMCC and several implementations with second
order. The difference between the versions is in the choice of the
parameters. As mentioned in Section~\pageref{sec:hllemcc}, for the standard
HLLEMCC, the parameter~\(\varepsilon\) in equations~\eqref{eq:119}
and~\eqref{eq:120} is chosen as~\(\varepsilon = 0.01\).  Here we also show
results for~\(\varepsilon = 0.005\),~\(\varepsilon = 0.00125\), and
for~\(\varepsilon = 10^{-6}\).  For the latter, the results are almost
indistinguishable from the pure second order HLLEM\@. All in all, the results
suggest to leave the parameters unchanged and stay with the same parameters as
in the standard version for first order. The gain in stability of the 1d shock
position and the loss in shear viscosity are just in balance. For the steady
shock test in Figure~\ref{fig:steadyccord2}, the situation slightly
improves. But it is still recommended to use HLLEMCC with the set of
parameters given in Section~\ref{sec:hllemcc}.


\subsubsection{Shear viscosity and the resolution of physical carbuncles}
\label{sec:shear-visc-resol}

In Section~\ref{sec:contribution-elling} we discussed the findings of
Elling~\cite{elling_carbuncle_2009} on physical carbuncles. This lead
us to the Elling test case as described in
Section~\ref{sec:elling-test} which allows us to test the numerical
flux functions for their ability to resolve these physical carbuncles
correctly. 
\begin{figure}
  \centering
  \includegraphics[width=\linewidth]{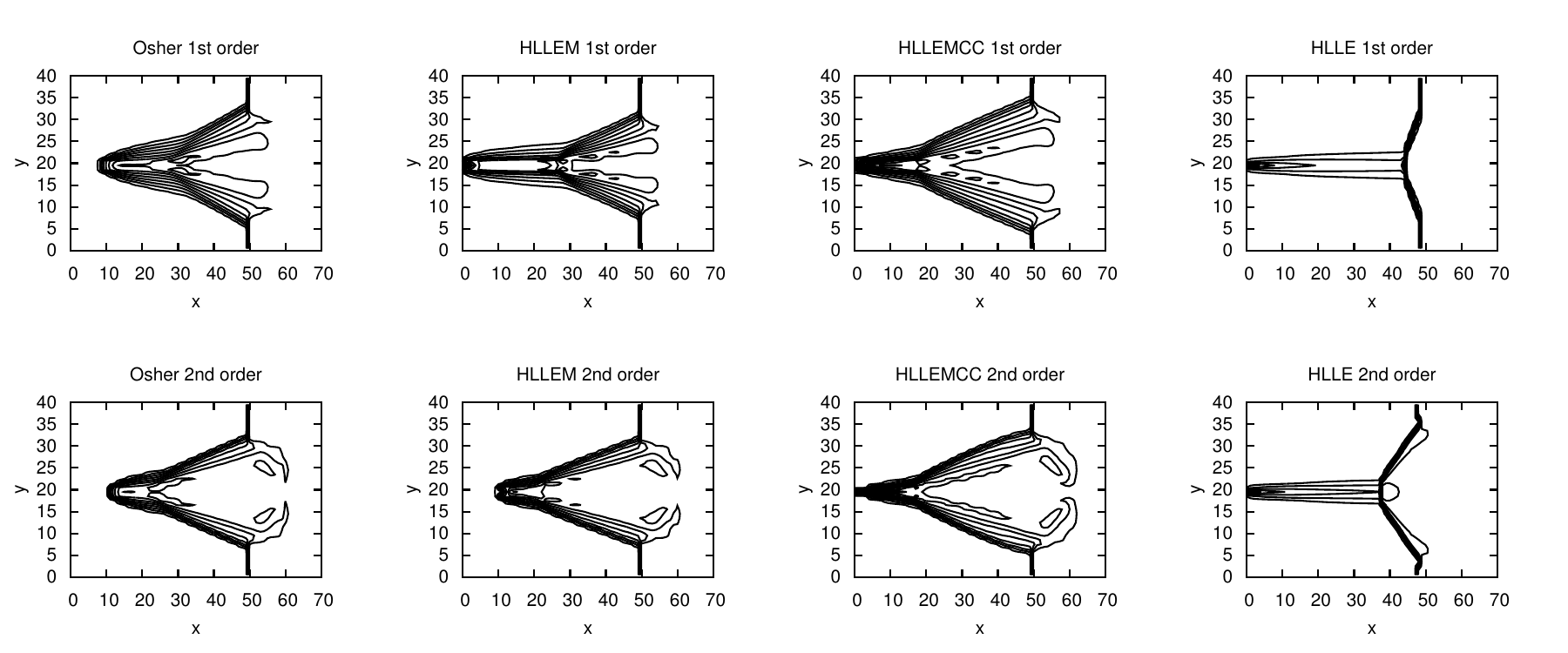}
  \caption{Elling test for gas dynamics at time~\(t=100\);
    comparison of different solvers; entropy shown. }
  \label{fig:ell15}
\end{figure}
As we can see from Figure~\ref{fig:ell15}, the numerical results with Osher,
HLLEM, and HLLEMCC are rather similar even with the first order scheme. The
HLLE scheme, which suppresses the carbuncle by a severe amount of
numerical shear viscosity, tries to prevent even the physical
carbuncle, not only in the first order, but also in the second order
computation. Since the test case is a model for shock boundary-layer and other
shock vortex interactions, we note that schemes which rely on the HLLE flux,
even when it is only applied locally in the vicinity of strong shocks, might
destroy some physical features of the flow. Thus, codes which are based on a
switch between complete and incomplete Riemann solvers depending on the
distance to the next strong shock should be carefully tested with the Elling
test before applying them to more complex flow problems. If the physical
carbuncle is not properly reproduced, the switching mechanism has to be
reworked.

\subsection{Influence of viscosity on entropy waves}
\label{sec:infl-visc-entr}

Some authors consider the carbuncle a result of the treatment of
mass transport and entropy waves~\cite{ausm-orig} within the Riemann
solver. The hunt for
entropy consistent Riemann solvers, e.\,g., is at least partially
driven by that idea. And indeed, for some schemes this causes
problems, e.\,g.\ for Flux Vector Splitting (FVS) schemes, which may
loose positivity by exactly resolving entropy
waves~\cite{gressier-positive}. But there is no strict proof, not even
the proof by Liou and Steffen in~\cite{ausm-orig}, for a connection
between the resolution of entropy waves and the carbuncle.

From our considerations in Section~\ref{sec:infl-one-dimens}, we know
that the instability of the 1d shock position causes a new Riemann
problem perpendicular to the original shock, which includes all types
of waves (for gas dynamics also entropy waves). Thus, although the
carbuncle occurs also in shallow water, where there are no entropy
waves, we can conclude that there is a connection between carbuncle
and entropy waves. The question we
have to answer is: Which type of waves has the stronger impact on the
stability of discrete shock profiles: shear waves or entropy waves?

A good means to answer that question is the HLLEMCC solver, which
allows to apply the carbuncle correction to both types of linearly
degenerate waves separately. 
\begin{figure}
  \centering
  \includegraphics[width=\linewidth]{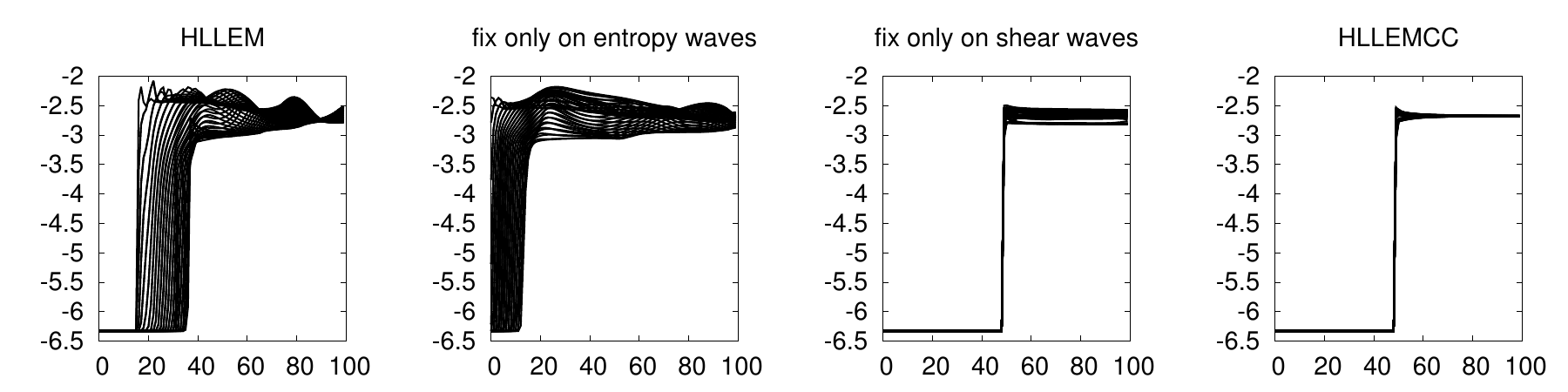}
  \caption{Steady shock in gas dynamics; comparison of HLLEM without
    carbuncle correction, correction on entropy waves, and full
    HLLEMCC at~\(t=1000\); entropy shown.}
  \label{fig:viscstead}
\end{figure}
\begin{figure}
  \centering
  \includegraphics[width=\linewidth]{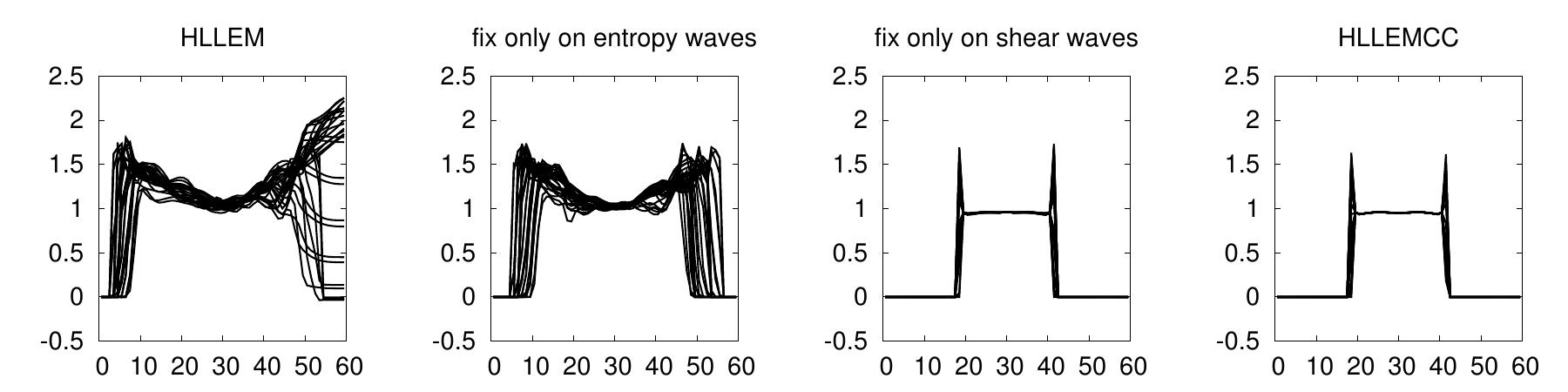}
  \caption{Colliding flow in gas dynamics; comparison of HLLEM without
    carbuncle correction, correction on entropy waves, and full
    HLLEMCC at~\(t=20\); entropy shown.}
  \label{fig:visccoll}
\end{figure}
In Figures~\ref{fig:viscstead} and~\ref{fig:visccoll}, we present
results for the steady shock and the colliding flow problem. We
compare pure HLLEM with three versions of HLLEMCC: correction only
applied to entropy waves, correction only applied to shear waves, full
HLLEMCC\@. From the numerical results we easily conclude that the
resolution of entropy waves contributes to the carbuncle, but the
contribution is small compared to the contribution by the shear
waves.

\section{Conclusions and directions for further research}
\label{sec:concl-direct-furth}

In this paper, we investigated the origin of the carbuncle
phenomenon. Guided by the theoretical results reviewed in
Section~\pageref{sec:review-theory}, we found a set of numerical test
cases which helped us to sort out the different issues involved in the
carbuncle. We observed that the Osher scheme by its special
construction of the numerical viscosity on shocks suppresses the
carbuncle to a certain extent. For steady grid-aligned shocks, we
could remodel this viscosity in the HLLEM scheme, confirming that it
is in fact the numerical viscosity on the shock which is responsible
for the stabilizing effect. We also showed how increasing the order
offers an alternative way to stabilize the shock position: introducing
more degrees of freedom for the scheme allows to remodel the
Rankine-Hugoniot condition at a captured shock. This stabilizing
mechanism even works when the mechanism of the Osher scheme is not
sufficient anymore: in the case of not perfectly-grid aligned shocks
like the bow shock in the blunt body problem.

Since the instability of the 1d shock position creates
vorticity along the shock, we also considered the influence of the
numerical viscosity on entropy and shear waves. We found that the
influence of the shear viscosity is much higher than that of the
viscosity on entropy waves. We also found
(cf.~Section~\pageref{sec:shear-visc-order}) that Riemann
solvers like HLLEMCC, which try to reduce the carbuncle from within
the Riemann solver and without too much loss of the resolution of
shear layers, should not be altered when used in a higher order
scheme. The gain in stability of the shock position is compensated by
the reduction of the shear viscosity. By employing the Elling test we
could show that incomplete Riemann solvers like HLLE not only prevent
non-physical carbuncles but also the physically induced breakdown of
the shock profile when it is hit by a vortex layer.

What is still lacking, is a deeper understanding of the amount of
numerical viscosity on shocks in order to stabilize the shock position
already for low order schemes. But this would be desirable when one
wants to combine the stabilizing mechanisms of Osher and HLLEMCC, in
which case the shear viscosity of HLLEMCC could be further
reduced. Furthermore, one would hope to be able to extend the
theoretical results reviewed in Section~\pageref{sec:review-theory} to
the case of higher order schemes. As we have seen, there is a
significant impact of the order of the scheme on the stability of
discrete shock profiles.

\section*{Acknowledgements}
\label{sec:achknowledgements}

This work was funded by Deutsche Forschungsgemeinschaft (DFG, KE
1420/3-1). 

\bibliographystyle{amsplain}
\bibliography{carb15}

\end{document}

%% file: elling-sketch-ink-bw.pdf_tex
\begingroup%
  \makeatletter%
  \providecommand\color[2][]{%
    \errmessage{(Inkscape) Color is used for the text in Inkscape, but the package 'color.sty' is not loaded}%
    \renewcommand\color[2][]{}%
  }%
  \providecommand\transparent[1]{%
    \errmessage{(Inkscape) Transparency is used (non-zero) for the text in Inkscape, but the package 'transparent.sty' is not loaded}%
    \renewcommand\transparent[1]{}%
  }%
  \providecommand\rotatebox[2]{#2}%
  \ifx\svgwidth\undefined%
    \setlength{\unitlength}{1139.21267393bp}%
    \ifx\svgscale\undefined%
      \relax%
    \else%
      \setlength{\unitlength}{\unitlength * \real{\svgscale}}%
    \fi%
  \else%
    \setlength{\unitlength}{\svgwidth}%
  \fi%
  \global\let\svgwidth\undefined%
  \global\let\svgscale\undefined%
  \makeatother%
  \begin{picture}(1,0.60165472)%
    \put(0,0){\includegraphics[width=\unitlength]{elling-sketch-ink-bw.pdf}}%
    \put(0.05188132,0.4501739){\color[rgb]{0,0,0}\makebox(0,0)[lb]{\smash{hypersonic inflow}}}%
    \put(0.05188132,0.25097492){\color[rgb]{0,0,0}\makebox(0,0)[lb]{\smash{filament with $\bm v=\bm 0$}}}%
    \put(0.64942559,0.35057441){\color[rgb]{0,0,0}\makebox(0,0)[lb]{\smash{post shock}}}%
    \put(0.64942559,0.30077466){\color[rgb]{0,0,0}\makebox(0,0)[lb]{\smash{region}}}%
    \put(0.64942559,0.40037416){\color[rgb]{0,0,0}\makebox(0,0)[lb]{\smash{subsonic}}}%
  \end{picture}%
\endgroup%

%% file: jumping-shock-15-ink.pdf_tex
\begingroup%
  \makeatletter%
  \providecommand\color[2][]{%
    \errmessage{(Inkscape) Color is used for the text in Inkscape, but the package 'color.sty' is not loaded}%
    \renewcommand\color[2][]{}%
  }%
  \providecommand\transparent[1]{%
    \errmessage{(Inkscape) Transparency is used (non-zero) for the text in Inkscape, but the package 'transparent.sty' is not loaded}%
    \renewcommand\transparent[1]{}%
  }%
  \providecommand\rotatebox[2]{#2}%
  \ifx\svgwidth\undefined%
    \setlength{\unitlength}{709.60050918bp}%
    \ifx\svgscale\undefined%
      \relax%
    \else%
      \setlength{\unitlength}{\unitlength * \real{\svgscale}}%
    \fi%
  \else%
    \setlength{\unitlength}{\svgwidth}%
  \fi%
  \global\let\svgwidth\undefined%
  \global\let\svgscale\undefined%
  \makeatother%
  \begin{picture}(1,0.54597374)%
    \put(0,0){\includegraphics[width=\unitlength]{jumping-shock-15-ink.pdf}}%
    \put(0.25264612,0.47763869){\color[rgb]{0,0,0}\makebox(0,0)[lb]{\smash{jump backwards}}}%
    \put(0.56270315,0.00938929){\color[rgb]{0,0,0}\makebox(0,0)[lb]{\smash{jump forward}}}%
    \put(0.60699701,0.52826024){\color[rgb]{0,0,0}\makebox(0,0)[lb]{\smash{original shock location}}}%
    \put(-0.00046167,0.18023704){\color[rgb]{0,0,0}\makebox(0,0)[lb]{\smash{$y$}}}%
    \put(0.18304147,0.00306159){\color[rgb]{0,0,0}\makebox(0,0)[lb]{\smash{$x$}}}%
    \put(0.75886168,0.41183066){\color[rgb]{0,0,0}\makebox(0,0)[lb]{\smash{new Riemann problem}}}%
    \put(0.57282746,0.30805647){\color[rgb]{0.62745098,0.62745098,0.62745098}\makebox(0,0)[lb]{\smash{$\bm v$}}}%
    \put(0.44247695,0.13594318){\color[rgb]{0.62745098,0.62745098,0.62745098}\makebox(0,0)[lb]{\smash{$\bm v$}}}%
  \end{picture}%
\endgroup%

%% file: shock-HO-bwink.pdf_tex
\begingroup%
  \makeatletter%
  \providecommand\color[2][]{%
    \errmessage{(Inkscape) Color is used for the text in Inkscape, but the package 'color.sty' is not loaded}%
    \renewcommand\color[2][]{}%
  }%
  \providecommand\transparent[1]{%
    \errmessage{(Inkscape) Transparency is used (non-zero) for the text in Inkscape, but the package 'transparent.sty' is not loaded}%
    \renewcommand\transparent[1]{}%
  }%
  \providecommand\rotatebox[2]{#2}%
  \ifx\svgwidth\undefined%
    \setlength{\unitlength}{1307.95635548bp}%
    \ifx\svgscale\undefined%
      \relax%
    \else%
      \setlength{\unitlength}{\unitlength * \real{\svgscale}}%
    \fi%
  \else%
    \setlength{\unitlength}{\svgwidth}%
  \fi%
  \global\let\svgwidth\undefined%
  \global\let\svgscale\undefined%
  \makeatother%
  \begin{picture}(1,0.519489)%
    \put(0,0){\includegraphics[width=\unitlength]{shock-HO-bwink.pdf}}%
    \put(0.18435697,0.1190257){\color[rgb]{0,0,0}\makebox(0,0)[lb]{\smash{\(q_l\)}}}%
    \put(0.8012473,0.37879311){\color[rgb]{0,0,0}\makebox(0,0)[lb]{\smash{\(q_r\)}}}%
    \put(0.51985551,0.1298837){\color[rgb]{0,0,0}\makebox(0,0)[lb]{\smash{\(q_m^-\)}}}%
    \put(0.60189994,0.12952238){\color[rgb]{0,0,0}\makebox(0,0)[lb]{\smash{\(q_m^+\)}}}%
    \put(0.95275538,0){\color[rgb]{0,0,0}\makebox(0,0)[lb]{\smash{x}}}%
    \put(0.26013392,0.45452424){\color[rgb]{0.50196078,0.50196078,0.50196078}\makebox(0,0)[lb]{\smash{1st order}}}%
    \put(0.26013392,0.41122967){\color[rgb]{0.30196078,0.30196078,0.30196078}\makebox(0,0)[lb]{\smash{2nd order}}}%
    \put(0.26013392,0.36798092){\color[rgb]{0,0,0}\makebox(0,0)[lb]{\smash{very high order}}}%
  \end{picture}%
\endgroup%